\documentclass[10pt,parskip=half,
toc=sectionentrywithdots,
bibliography=totocnumbered,
captions=tableheading,numbers=noendperiod]{scrartcl}

\usepackage[T1]{fontenc} 
\usepackage{mathpazo}
\usepackage{graphicx}
\usepackage[skip=3pt]{caption}
\usepackage{adjustbox} 
\usepackage[table]{xcolor} 
\usepackage{enumerate} 
\usepackage{amsmath} 
\usepackage{amssymb} 
\usepackage{textcomp} 
\AtBeginDocument{%
}
\usepackage{upquote} 
\usepackage{eurosym} 
\usepackage[mathletters]{ucs} 
\usepackage[utf8x]{inputenc} 
\usepackage{fancyvrb} 
\usepackage{grffile} 
\usepackage{hyperref}
\usepackage{longtable} 
\usepackage{booktabs}  
\usepackage[inline]{enumitem} 
\usepackage[normalem]{ulem} 

\usepackage{translations}
\usepackage{microtype} 
\usepackage{placeins} 
\usepackage{float}
\usepackage[colorinlistoftodos,obeyFinal,textwidth=.8in]{todonotes} 
\let\counterwithout\relax

\usepackage{chngcntr}
\usepackage[footsepline=0.25pt]{scrlayer-scrpage}

\usepackage[sort,super]{natbib}
\usepackage{doi}

    \usepackage{newfloat} 
    \DeclareFloatingEnvironment[
        fileext=frm,placement={!ht},
        within=section,name=Code]{codecell}
    \DeclareFloatingEnvironment[
        fileext=frm,placement={!ht},
        within=section,name=Text]{textcell}
    \DeclareFloatingEnvironment[
        fileext=frm,placement={!ht},
        within=section,name=Text]{errorcell}

    \usepackage{listings} 
    \usepackage[framemethod=tikz]{mdframed} 

\makeatletter
\def\PY@reset{\let\PY@it=\relax \let\PY@bf=\relax%
    \let\PY@ul=\relax \let\PY@tc=\relax%
    \let\PY@bc=\relax \let\PY@ff=\relax}
\def\PY@tok#1{\csname PY@tok@#1\endcsname}
\def\PY@toks#1+{\ifx\relax#1\empty\else%
    \PY@tok{#1}\expandafter\PY@toks\fi}
\def\PY@do#1{\PY@bc{\PY@tc{\PY@ul{%
    \PY@it{\PY@bf{\PY@ff{#1}}}}}}}
\def\PY#1#2{\PY@reset\PY@toks#1+\relax+\PY@do{#2}}

\expandafter\def\csname PY@tok@w\endcsname{\def\PY@tc##1{\textcolor[rgb]{0.73,0.73,0.73}{##1}}}
\expandafter\def\csname PY@tok@c\endcsname{\let\PY@it=\textit\def\PY@tc##1{\textcolor[rgb]{0.25,0.50,0.50}{##1}}}
\expandafter\def\csname PY@tok@cp\endcsname{\def\PY@tc##1{\textcolor[rgb]{0.74,0.48,0.00}{##1}}}
\expandafter\def\csname PY@tok@k\endcsname{\let\PY@bf=\textbf\def\PY@tc##1{\textcolor[rgb]{0.00,0.50,0.00}{##1}}}
\expandafter\def\csname PY@tok@kp\endcsname{\def\PY@tc##1{\textcolor[rgb]{0.00,0.50,0.00}{##1}}}
\expandafter\def\csname PY@tok@kt\endcsname{\def\PY@tc##1{\textcolor[rgb]{0.69,0.00,0.25}{##1}}}
\expandafter\def\csname PY@tok@o\endcsname{\def\PY@tc##1{\textcolor[rgb]{0.40,0.40,0.40}{##1}}}
\expandafter\def\csname PY@tok@ow\endcsname{\let\PY@bf=\textbf\def\PY@tc##1{\textcolor[rgb]{0.67,0.13,1.00}{##1}}}
\expandafter\def\csname PY@tok@nb\endcsname{\def\PY@tc##1{\textcolor[rgb]{0.00,0.50,0.00}{##1}}}
\expandafter\def\csname PY@tok@nf\endcsname{\def\PY@tc##1{\textcolor[rgb]{0.00,0.00,1.00}{##1}}}
\expandafter\def\csname PY@tok@nc\endcsname{\let\PY@bf=\textbf\def\PY@tc##1{\textcolor[rgb]{0.00,0.00,1.00}{##1}}}
\expandafter\def\csname PY@tok@nn\endcsname{\let\PY@bf=\textbf\def\PY@tc##1{\textcolor[rgb]{0.00,0.00,1.00}{##1}}}
\expandafter\def\csname PY@tok@ne\endcsname{\let\PY@bf=\textbf\def\PY@tc##1{\textcolor[rgb]{0.82,0.25,0.23}{##1}}}
\expandafter\def\csname PY@tok@nv\endcsname{\def\PY@tc##1{\textcolor[rgb]{0.10,0.09,0.49}{##1}}}
\expandafter\def\csname PY@tok@no\endcsname{\def\PY@tc##1{\textcolor[rgb]{0.53,0.00,0.00}{##1}}}
\expandafter\def\csname PY@tok@nl\endcsname{\def\PY@tc##1{\textcolor[rgb]{0.63,0.63,0.00}{##1}}}
\expandafter\def\csname PY@tok@ni\endcsname{\let\PY@bf=\textbf\def\PY@tc##1{\textcolor[rgb]{0.60,0.60,0.60}{##1}}}
\expandafter\def\csname PY@tok@na\endcsname{\def\PY@tc##1{\textcolor[rgb]{0.49,0.56,0.16}{##1}}}
\expandafter\def\csname PY@tok@nt\endcsname{\let\PY@bf=\textbf\def\PY@tc##1{\textcolor[rgb]{0.00,0.50,0.00}{##1}}}
\expandafter\def\csname PY@tok@nd\endcsname{\def\PY@tc##1{\textcolor[rgb]{0.67,0.13,1.00}{##1}}}
\expandafter\def\csname PY@tok@s\endcsname{\def\PY@tc##1{\textcolor[rgb]{0.73,0.13,0.13}{##1}}}
\expandafter\def\csname PY@tok@sd\endcsname{\let\PY@it=\textit\def\PY@tc##1{\textcolor[rgb]{0.73,0.13,0.13}{##1}}}
\expandafter\def\csname PY@tok@si\endcsname{\let\PY@bf=\textbf\def\PY@tc##1{\textcolor[rgb]{0.73,0.40,0.53}{##1}}}
\expandafter\def\csname PY@tok@se\endcsname{\let\PY@bf=\textbf\def\PY@tc##1{\textcolor[rgb]{0.73,0.40,0.13}{##1}}}
\expandafter\def\csname PY@tok@sr\endcsname{\def\PY@tc##1{\textcolor[rgb]{0.73,0.40,0.53}{##1}}}
\expandafter\def\csname PY@tok@ss\endcsname{\def\PY@tc##1{\textcolor[rgb]{0.10,0.09,0.49}{##1}}}
\expandafter\def\csname PY@tok@sx\endcsname{\def\PY@tc##1{\textcolor[rgb]{0.00,0.50,0.00}{##1}}}
\expandafter\def\csname PY@tok@m\endcsname{\def\PY@tc##1{\textcolor[rgb]{0.40,0.40,0.40}{##1}}}
\expandafter\def\csname PY@tok@gh\endcsname{\let\PY@bf=\textbf\def\PY@tc##1{\textcolor[rgb]{0.00,0.00,0.50}{##1}}}
\expandafter\def\csname PY@tok@gu\endcsname{\let\PY@bf=\textbf\def\PY@tc##1{\textcolor[rgb]{0.50,0.00,0.50}{##1}}}
\expandafter\def\csname PY@tok@gd\endcsname{\def\PY@tc##1{\textcolor[rgb]{0.63,0.00,0.00}{##1}}}
\expandafter\def\csname PY@tok@gi\endcsname{\def\PY@tc##1{\textcolor[rgb]{0.00,0.63,0.00}{##1}}}
\expandafter\def\csname PY@tok@gr\endcsname{\def\PY@tc##1{\textcolor[rgb]{1.00,0.00,0.00}{##1}}}
\expandafter\def\csname PY@tok@ge\endcsname{\let\PY@it=\textit}
\expandafter\def\csname PY@tok@gs\endcsname{\let\PY@bf=\textbf}
\expandafter\def\csname PY@tok@gp\endcsname{\let\PY@bf=\textbf\def\PY@tc##1{\textcolor[rgb]{0.00,0.00,0.50}{##1}}}
\expandafter\def\csname PY@tok@go\endcsname{\def\PY@tc##1{\textcolor[rgb]{0.53,0.53,0.53}{##1}}}
\expandafter\def\csname PY@tok@gt\endcsname{\def\PY@tc##1{\textcolor[rgb]{0.00,0.27,0.87}{##1}}}
\expandafter\def\csname PY@tok@err\endcsname{\def\PY@bc##1{\setlength{\fboxsep}{0pt}\fcolorbox[rgb]{1.00,0.00,0.00}{1,1,1}{\strut ##1}}}
\expandafter\def\csname PY@tok@kc\endcsname{\let\PY@bf=\textbf\def\PY@tc##1{\textcolor[rgb]{0.00,0.50,0.00}{##1}}}
\expandafter\def\csname PY@tok@kd\endcsname{\let\PY@bf=\textbf\def\PY@tc##1{\textcolor[rgb]{0.00,0.50,0.00}{##1}}}
\expandafter\def\csname PY@tok@kn\endcsname{\let\PY@bf=\textbf\def\PY@tc##1{\textcolor[rgb]{0.00,0.50,0.00}{##1}}}
\expandafter\def\csname PY@tok@kr\endcsname{\let\PY@bf=\textbf\def\PY@tc##1{\textcolor[rgb]{0.00,0.50,0.00}{##1}}}
\expandafter\def\csname PY@tok@bp\endcsname{\def\PY@tc##1{\textcolor[rgb]{0.00,0.50,0.00}{##1}}}
\expandafter\def\csname PY@tok@fm\endcsname{\def\PY@tc##1{\textcolor[rgb]{0.00,0.00,1.00}{##1}}}
\expandafter\def\csname PY@tok@vc\endcsname{\def\PY@tc##1{\textcolor[rgb]{0.10,0.09,0.49}{##1}}}
\expandafter\def\csname PY@tok@vg\endcsname{\def\PY@tc##1{\textcolor[rgb]{0.10,0.09,0.49}{##1}}}
\expandafter\def\csname PY@tok@vi\endcsname{\def\PY@tc##1{\textcolor[rgb]{0.10,0.09,0.49}{##1}}}
\expandafter\def\csname PY@tok@vm\endcsname{\def\PY@tc##1{\textcolor[rgb]{0.10,0.09,0.49}{##1}}}
\expandafter\def\csname PY@tok@sa\endcsname{\def\PY@tc##1{\textcolor[rgb]{0.73,0.13,0.13}{##1}}}
\expandafter\def\csname PY@tok@sb\endcsname{\def\PY@tc##1{\textcolor[rgb]{0.73,0.13,0.13}{##1}}}
\expandafter\def\csname PY@tok@sc\endcsname{\def\PY@tc##1{\textcolor[rgb]{0.73,0.13,0.13}{##1}}}
\expandafter\def\csname PY@tok@dl\endcsname{\def\PY@tc##1{\textcolor[rgb]{0.73,0.13,0.13}{##1}}}
\expandafter\def\csname PY@tok@s2\endcsname{\def\PY@tc##1{\textcolor[rgb]{0.73,0.13,0.13}{##1}}}
\expandafter\def\csname PY@tok@sh\endcsname{\def\PY@tc##1{\textcolor[rgb]{0.73,0.13,0.13}{##1}}}
\expandafter\def\csname PY@tok@s1\endcsname{\def\PY@tc##1{\textcolor[rgb]{0.73,0.13,0.13}{##1}}}
\expandafter\def\csname PY@tok@mb\endcsname{\def\PY@tc##1{\textcolor[rgb]{0.40,0.40,0.40}{##1}}}
\expandafter\def\csname PY@tok@mf\endcsname{\def\PY@tc##1{\textcolor[rgb]{0.40,0.40,0.40}{##1}}}
\expandafter\def\csname PY@tok@mh\endcsname{\def\PY@tc##1{\textcolor[rgb]{0.40,0.40,0.40}{##1}}}
\expandafter\def\csname PY@tok@mi\endcsname{\def\PY@tc##1{\textcolor[rgb]{0.40,0.40,0.40}{##1}}}
\expandafter\def\csname PY@tok@il\endcsname{\def\PY@tc##1{\textcolor[rgb]{0.40,0.40,0.40}{##1}}}
\expandafter\def\csname PY@tok@mo\endcsname{\def\PY@tc##1{\textcolor[rgb]{0.40,0.40,0.40}{##1}}}
\expandafter\def\csname PY@tok@ch\endcsname{\let\PY@it=\textit\def\PY@tc##1{\textcolor[rgb]{0.25,0.50,0.50}{##1}}}
\expandafter\def\csname PY@tok@cm\endcsname{\let\PY@it=\textit\def\PY@tc##1{\textcolor[rgb]{0.25,0.50,0.50}{##1}}}
\expandafter\def\csname PY@tok@cpf\endcsname{\let\PY@it=\textit\def\PY@tc##1{\textcolor[rgb]{0.25,0.50,0.50}{##1}}}
\expandafter\def\csname PY@tok@c1\endcsname{\let\PY@it=\textit\def\PY@tc##1{\textcolor[rgb]{0.25,0.50,0.50}{##1}}}
\expandafter\def\csname PY@tok@cs\endcsname{\let\PY@it=\textit\def\PY@tc##1{\textcolor[rgb]{0.25,0.50,0.50}{##1}}}

\makeatother

\definecolor{ansi-black}{HTML}{3E424D}
\definecolor{ansi-black-intense}{HTML}{282C36}
\definecolor{ansi-red}{HTML}{E75C58}
\definecolor{ansi-red-intense}{HTML}{B22B31}
\definecolor{ansi-green}{HTML}{00A250}
\definecolor{ansi-green-intense}{HTML}{007427}
\definecolor{ansi-yellow}{HTML}{DDB62B}
\definecolor{ansi-yellow-intense}{HTML}{B27D12}
\definecolor{ansi-blue}{HTML}{208FFB}
\definecolor{ansi-blue-intense}{HTML}{0065CA}
\definecolor{ansi-magenta}{HTML}{D160C4}
\definecolor{ansi-magenta-intense}{HTML}{A03196}
\definecolor{ansi-cyan}{HTML}{60C6C8}
\definecolor{ansi-cyan-intense}{HTML}{258F8F}
\definecolor{ansi-white}{HTML}{C5C1B4}
\definecolor{ansi-white-intense}{HTML}{A1A6B2}

\providecommand{\tightlist}{%
  \setlength{\itemsep}{0pt}\setlength{\parskip}{0pt}}
\DefineVerbatimEnvironment{Highlighting}{Verbatim}{commandchars=\\\{\}}

\definecolor{urlcolor}{rgb}{0,.145,.698}
\definecolor{linkcolor}{rgb}{.71,0.21,0.01}
\definecolor{citecolor}{rgb}{.12,.54,.11}

\DeclareTranslationFallback{Author}{Author}
\DeclareTranslation{Portuges}{Author}{Autor}

\DeclareTranslationFallback{List of Codes}{List of Codes}
\DeclareTranslation{Catalan}{List of Codes}{Llista de Codis}
\DeclareTranslation{Danish}{List of Codes}{Liste over Koder}
\DeclareTranslation{German}{List of Codes}{Liste der Codes}
\DeclareTranslation{Spanish}{List of Codes}{Lista de C\'{o}digos}
\DeclareTranslation{French}{List of Codes}{Liste des Codes}
\DeclareTranslation{Italian}{List of Codes}{Elenco dei Codici}
\DeclareTranslation{Dutch}{List of Codes}{Lijst van Codes}
\DeclareTranslation{Portuges}{List of Codes}{Lista de C\'{o}digos}

\DeclareTranslationFallback{Supervisors}{Supervisors}
\DeclareTranslation{Catalan}{Supervisors}{Supervisors}
\DeclareTranslation{Danish}{Supervisors}{Vejledere}
\DeclareTranslation{German}{Supervisors}{Vorgesetzten}
\DeclareTranslation{Spanish}{Supervisors}{Supervisores}
\DeclareTranslation{French}{Supervisors}{Superviseurs}
\DeclareTranslation{Italian}{Supervisors}{Le autorit\`{a} di vigilanza}
\DeclareTranslation{Dutch}{Supervisors}{supervisors}
\DeclareTranslation{Portuguese}{Supervisors}{Supervisores}

\definecolor{codegreen}{rgb}{0,0.6,0}
\definecolor{codegray}{rgb}{0.5,0.5,0.5}
\definecolor{codepurple}{rgb}{0.58,0,0.82}
\definecolor{backcolour}{rgb}{0.95,0.95,0.95}

\lstdefinestyle{mystyle}{
    commentstyle=\color{codegreen},
    keywordstyle=\color{magenta},
    numberstyle=\tiny\color{codegray},
    stringstyle=\color{codepurple},
    basicstyle=\ttfamily,
    breakatwhitespace=false,
    keepspaces=true,
    numbers=none,
    numbersep=10pt,
    showspaces=false,
    showstringspaces=false,
    showtabs=false,
    tabsize=2,
    breaklines=true,
    literate={\-}{}{0\discretionary{-}{}{-}},
  postbreak=\mbox{\textcolor{red}{$\hookrightarrow$}\space},
}

\surroundwithmdframed[
  hidealllines=true,
  backgroundcolor=backcolour,
  innerleftmargin=0pt,
  innerrightmargin=0pt,
  innertopmargin=0pt,
  innerbottommargin=0pt]{lstlisting}

\usepackage{geometry}
\addtokomafont{section}{\clearpage}

\hypersetup{
    breaklinks=true,  
    colorlinks=true,
    urlcolor=urlcolor,
    linkcolor=linkcolor,
    citecolor=citecolor,
    }

\makeatletter
\AtBeginDocument{%
    \expandafter\renewcommand\expandafter\subsection\expandafter
    {\expandafter\@fb@secFB\subsection}%
    \newcommand\@fb@secFB{\FloatBarrier
    \gdef\@fb@afterHHook{\@fb@topbarrier \gdef\@fb@afterHHook{}}}%
    \g@addto@macro\@afterheading{\@fb@afterHHook}%
    \gdef\@fb@afterHHook{}%
}
\makeatother

\counterwithout{figure}{section}
\counterwithout{table}{section}
\counterwithout{equation}{section}
\makeatletter
\@addtoreset{table}{section}
\@addtoreset{figure}{section}
\@addtoreset{equation}{section}
\makeatother

    \makeatletter
        \providecommand*\setfloatlocations[2]{\@namedef{fps@#1}{#2}}
    \makeatother

\ModifyLayer[addvoffset=.6ex]{scrheadings.foot.odd}
\ModifyLayer[addvoffset=.6ex]{scrheadings.foot.even}
\ModifyLayer[addvoffset=.6ex]{scrheadings.foot.oneside}
\ModifyLayer[addvoffset=.6ex]{plain.scrheadings.foot.odd}
\ModifyLayer[addvoffset=.6ex]{plain.scrheadings.foot.even}
\ModifyLayer[addvoffset=.6ex]{plain.scrheadings.foot.oneside}
\clearscrheadfoot{}
\ifoot{\leftmark}

\ofoot{\pagemark}
\usepackage{cleveref}
\creflabelformat{equation}{#2#1#3}

\crefname{codecell}{code}{codes}
\Crefname{codecell}{code}{codes}
\crefname{textcell}{text}{texts}
\Crefname{textcell}{text}{texts}
\crefname{errorcell}{error}{errors}
\Crefname{errorcell}{error}{errors}

\begin{document}

    \begin{titlepage}

  \begin{center}

  \vspace*{1cm}

  \Huge\textbf{Fast Cubic Spline Interpolation}

  \vspace{0.5cm}

  \vspace{1.5cm}

  \begin{minipage}{0.8\textwidth}
    \begin{center}
    \begin{minipage}{0.39\textwidth}
    \begin{flushleft} \Large
    \emph{\GetTranslation{Author}:}\\Haysn Hornbeck\\\href{mailto:hhornbec@ucalgary.ca}{hhornbec@ucalgary.ca}
    \end{flushleft}
    \end{minipage}
    \hspace{\fill}
    \begin{minipage}{0.39\textwidth}
    \begin{flushright} \Large
    \end{flushright}
    \end{minipage}
    \end{center}
  \end{minipage}

  \vfill

  \begin{minipage}{0.8\textwidth}
  \begin{center}
  \end{center}
  \end{minipage}

  \vspace{0.8cm}
      \LARGE{University of Calgary}\\

  \vspace{0.4cm}

  \today

  \end{center}
  \end{titlepage}

    \begingroup
    \let\cleardoublepage\relax
    \let\clearpage\relax\tableofcontents\listoffigures\listoftables\listof{codecell}{\GetTranslation{List of Codes}}
    \endgroup

\hypertarget{abstract}{%
\subsection{Abstract}\label{abstract}}

The \emph{Numerical Recipes} series of books are a useful resource, but
all the algorithms they contain cannot be used within open-source
projects. In this paper we develop drop-in alternatives to the two
algorithms they present for cubic spline interpolation, showing as much
of our work as possible to allow for replication or criticsm. The output
of the new algorithms is compared to the old, and found to be no
different within the limits imposed by floating-point precision.
Benchmarks of all these algorithms, plus variations which may run faster
in certain instances, are performed. In general, all these algorithms
have approximately the same execution time when interpolating curves
with few control points on feature-rich Intel processors; as the number
of control points increases or processor features are removed, the new
algorithms become consistently faster than the old. Exceptions to that
generalization are explored to create implementation guidelines, such as
when to expect division to be faster than multiplication.

\hypertarget{introduction}{%
\section{Introduction}\label{introduction}}

Since first being released in 1986, the \emph{Numerical Recipies} series
has sold over a quarter-million copies, according to its
authors,\cite{noauthor_numerical_2015} and been cited tens of thousands
of times, according to Google Scholar. Several editions have been
published, the most recent in 2007, with the computer code in this
series translated into C, Fortran, Pascal, BASIC, and C++. This level of
popularity is unusual for a thousand-page technical publication that
consists of hundreds of small algorithms intended for use in engineering
and scientific computing. The series' appeal can be explained by the
breadth of algorithms it includes, a desire to describe ``what is inside
the black box'',\cite{press1997numerical} as well as an informal writing
style.

\begin{quote}
The two Numerical Recipes books are marvellous. The principal book, The
Art of Scientific Computing, contains program listings for almost every
conceivable requirement, and it also contains a well written discussion
of the algorithms and the numerical methods involved. The Example Book
provides a complete driving program, with helpful notes, for nearly all
the routines in the principal book.\cite{Press_2003}
\end{quote}

\begin{quote}
The authors express their opinions on methods and their usage and offer
valuable insights in a highly readable and personable prose. For the
novice or less mathematically inclined user, this guidance makes it
possible to arrive at solutions using the computer routines more or less
as black boxes. The more seasoned user can easily tailor the routines to
his specific needs, assisted by the clear and consistent code, the
inclusion of utility routines, and the meticulous documentation
throughout.\cite{doi:10.1111/j.1539-6924.1989.tb01007.x}
\end{quote}

\begin{quote}
The two central characteristics of this book that I find the most
compelling are its informality and the readiness of the authors to go
out on a limb and present their own opinions. \ldots{} Formality and a
``close'' style offer a warm refuge from a more daring and perceptive
use of the language. This should not be so, as has been repeatedly
demonstrated in a few excellent books (e.g.~Gilbert Strang's
\emph{Introduction to Applied Mathematics})-it is possible, without
compromising mathematical precision, to write well and comprehensibly
(and even with a sense of humour). This is also the impression with
\emph{Numerical Recipes} and, reading it, I repeatedly felt that the
authors are simply enjoying the task of explaining their subject matter.
This enjoyment will be shared by the readers.\cite{iserles_1989}
\end{quote}

The \emph{Numerical Recipes} series has not escaped criticism, however.
The authors created a web page devoted to countering rumours that the
code within their book was full of
mistakes.\cite{noauthor_numerical_2015} Other reviewers have argued the
text contains inefficient, outdated, or inaccurate algorithms. Pavel
Holoborodko tested a number of algorithms and libraries for calculating
modified Bessel functions, and according to Jutta Degener found that
``tests reveal that \emph{Numerical Recipes} algorithms delivers the
lowest accuracy in many cases!''\cite{degener_2016} Even the format and
style which makes the series appealing has been a source of critique.

\begin{quote}
In order to present a great many topics, the authors devote little space
to any one method. The treatment of each method is adequate to
appreciate the basic idea, and references are provided to the
literature. Generally the treatment is too superficial for a textbook.
Further, the lack of examples, illustrative computations, and exercises
make the book unsuitable for the classroom. \ldots{} The codes
themselves are of better than average quality for a survey book.
However, they are far from being mathematical software. Even in the
cases of a high quality code taken from the literature, the
documentation expected of mathematical software is absent. Although the
authors repeatedly express their distaste for ``black boxes,'' they do
refer the reader to such codes in a number of instances. With few
exceptions, the reader would be well advised to turn to reputable
sources of mathematical software rather than to the codes given in this
book. The advice offered does not always correspond to the methods
advocated by leading practitioners and implemented in leading
libraries.\cite{doi:10.1080/00029890.1987.12000737}
\end{quote}

The critique of \emph{Numerical Recipes} we are most concerned with is
the license the programming code is released under. One of two licenses
may apply, depending on whether access was purchased as an individual or
as an institution.

\begin{quote}
By purchasing this disk or code download, you acquire a Numerical
Recipes Personal Single-User License. This license lets you personally
use Numerical Recipes code (``the code'') on any number of computers,
but only one computer at a time. You are not permitted to allow anyone
else to access or use the code. You may, under this license, transfer
precompiled, executable applications incorporating the code to other,
unlicensed, persons, providing that (i) the application is noncommercial
(e.g., does not involve the selling or licensing of the application for
a fee or its use in developing commercial products or services), and
(ii) the application was first developed, compiled, and successfully run
by you, and (iii) the code is bound into the application in such a
manner that it cannot be accessed as individual routines and cannot
practicably be unbound and used in other programs. That is, under this
license, your application user must not be able to use Numerical Recipes
code as part of a program library or ``mix and match''
workbench.\cite{about_numerical_recipes_2015}
\end{quote}

The institutional license allows computers with a range of IP addresses
to share \emph{Numerical Recipes} programming code among themselves, but
not to any computer outside that range. The license explicitly forbids
the use of Network Address Translation or proxies to circumvent this
restriction. Executables may be shared and sold outside this range,
provided they were compiled within this range and \emph{Numerical
Recipes}' source code ``cannot practicably be unbound and used in other
programs.'' Both licenses are unusually restrictive and impractical.

\begin{quote}
Free sharing of data after publication is a requirement in science (read
the \emph{Astrophysical Journal} policy if you don't believe me).
Algorithms and code are not data, but the tradition of sharing them
helps the sciences to develop. The net effect of the NR license on
science is to discourage people from helping others by sharing their
work. I personally have a body of useful code that I want to distribute,
but haven't devoted the time to extracting it from the death grip of the
NR license, because in a publish-or-perish world, I can't justify
spending the time when I could be writing a paper
instead.\cite{benjamin_weiner_boycott_2006}
\end{quote}

Both licenses are also incompatible with open-source software, where the
source code must be shared with whoever asks for it. These restrictions
were a major obstacle for us, when we found ourselves in need of a cubic
B-spline interpolation routine. The version published in \emph{Numerical
Recipes} was perfect for our needs, and yet because the code would be
used in a scientific publication and redistributed under an open-source
license we found ourselves unable to use it. The available alternatives
were either part of an existing library or poorly documented.

Our goal for this paper is to create a functional equivalent of the
cubic B-spline code in \emph{Numerical Recipes}. By deriving this code
from the relevant mathematics, without reference to the original code,
we are free to release our code under the license of our chosing.

\begin{quote}
Dedicating works to the public domain is difficult if not impossible for
those wanting to contribute their works for public use before applicable
copyright or database protection terms expire. Few if any jurisdictions
have a process for doing so easily and reliably. Laws vary from
jurisdiction to jurisdiction as to what rights are automatically granted
and how and when they expire or may be voluntarily relinquished. More
challenging yet, many legal systems effectively prohibit any attempt by
these owners to surrender rights automatically conferred by law,
particularly moral rights, even when the author wishing to do so is well
informed and resolute about doing so and contributing their work to the
public domain.
\end{quote}

\begin{quote}
CC0 helps solve this problem by giving creators a way to waive all their
copyright and related rights in their works to the fullest extent
allowed by law. CC0 is a universal instrument that is not adapted to the
laws of any particular legal jurisdiction, similar to many open source
software licenses. And while no tool, not even CC0, can guarantee a
complete relinquishment of all copyright and database rights in every
jurisdiction, we believe it provides the best and most complete
alternative for contributing a work to the public domain given the many
complex and diverse copyright and database systems around the
world.\cite{cc0_explain}
\end{quote}

We thus release the two algorithms necessary for cubic B-spline
interpolation under the Creative Commons Zero license.\cite{cc0_legal}
Unless stated otherwise, the other source code relating to the
derivation of those two routines is released under the Creative Commons
Attribution-ShareAlike 4.0 International license.\cite{ccbyasa}. To
conserve space, that source code is not embedded within this document
and can instead be obtained online.\cite{fast_cubic_source}

Since the author has purchased one edition of the book, we are allowed
access to the original code under the \emph{Numerical Recipes} Personal
Single-User License. We will take advantage of this to also compare the
accuracy and efficiency of our replacement routines to those of the
originals. These comparisons should be helpful to others interested in a
replacing the original \emph{Numerical Recipes} code.

\hypertarget{fast-interpolation}{%
\section{Fast Interpolation}\label{fast-interpolation}}

We can think of B-spline interpolation as a recreation of an unknown
function based on sparse inputs. Each knot, \(x_j\), is fed into the
function and generates an output value, \(y_j = f(x_j)\). The usual
approach to cubic B-spline interpolation requires four knots and values
to be referenced, in order to guarantee the proper continuity.
\emph{Numerical Recipes} points out there is another method which relies
on two knots and values, plus two second derivatives.

\begin{align}
y &= A y_j + B y_{j+1} + C y''_j + D y''_{j+1}, \label{eqn:nr_formula}\\
A &= \frac{ x_{j+1} - x }{ x_{j+1} - x_j }, \\
B &= \frac{ x - x_{j} }{ x_{j+1} - x_j }, \\
C &= \frac 1 6 (A^3 - A)( x_{j+1} - x_j )^2, \\
D &= \frac 1 6 (B^3 - B)( x_{j+1} - x_j )^2,
\end{align}

where \(x\) is the location we wish to interpolate at, \(x_j\) is the
\(j\)th knot, \(y''_j\) is the second derivative at \(x_j\),
\(x_{j+1} \ge x \ge x_j\), and \(x_{j+1} > x_j\). \emph{Numerical
Recipes} do not go into detail on how they derived this equation, but
Arne Morten Kvarving fills in some of the details in their lecture
notes.\cite{kvarving_natural_2008}

Some analysis reveals that a number of the \(( x_{j+1} - x_j )\) terms
in \(C\) and \(D\) will cancel out. Manually canceling these values out
may reduce the number of operations necessary. \texttt{Sympy} can
automate much of this work for us.\cite{meurer2017sympy} We begin with
the original equation.

\begin{lstlisting}[language=Python,xleftmargin=20pt,xrightmargin=5pt,belowskip=5pt,aboveskip=5pt]
import sympy as sp

def peq( math ):
    txt = str(math)
    latex = sp.latex(math)
    display({'text/latex': "$$" + latex + "$$",
        'text/txt': txt
            }, raw=True)

knots = sp.symbols('x_{j-3} x_{j-2} x_{j-1} x_{j} x_{j+1} x_{j+2} x_{j+3} x_{j+4}', real=True)
values = sp.symbols('y_{j-1} y_{j} y_{j+1} y_{j+2}', real=True)
second_values = sp.symbols("y''_{j-1} y''_{j} y''_{j+1} y''_{j+2}", real=True)
x, y, ypp, A, B, C, D = sp.symbols("x y y'' A B C D", real=True)

def xj(n):
    return knots[n+3]

def yj(n):
    return values[n+1]

def yppj(n):
    return second_values[n+1]

def compare_math( a, b ):
    temp = sp.simplify( a - b )
    if temp == 0:
        display( {'text/latex': "The two statements are equivalent.",
                  'text/txt': "The two statements are equivalent."}, raw=True)
    else:
        peq( temp )

nr_interpolate = sp.Eq( y, A*yj(0) + B*yj(1) + C*yppj(0) + D*yppj(1) )
peq( nr_interpolate )
\end{lstlisting}

\begin{align}\label{eq:nr_dupe}
y = A y_{j} + B y_{j+1} + C y''_{j} + D y''_{j+1}
\end{align}

Next, we perform the substitutions, being careful of the order of
substitution.

\begin{lstlisting}[language=Python,xleftmargin=20pt,xrightmargin=5pt,belowskip=5pt,aboveskip=5pt]
complex_y = sp.Eq( y, nr_interpolate.rhs.subs( [(C, (A**3 - A)*( (xj(1) - xj(0))**2 )/6), (D, (B**3 - B)*( (xj(1) - xj(0))**2 )/6), \
                    (A, (xj(1) - x)/(xj(1) - xj(0))), (B, (x - xj(0))/(xj(1) - xj(0)))] ) )
\end{lstlisting}

Finally, we ask \texttt{Sympy} to perform the cancellations.

\begin{lstlisting}[language=Python,xleftmargin=20pt,xrightmargin=5pt,belowskip=5pt,aboveskip=5pt]
simpler_y = sp.Eq( y, sp.simplify( complex_y.rhs ) )
peq( simpler_y )
\end{lstlisting}

$$y = \frac{\frac{y''_{j+1} \left(\left(- x + x_{j}\right) \left(x_{j+1} - x_{j}\right)^{2} + \left(x - x_{j}\right)^{3}\right)}{6} - \frac{y''_{j} \left(\left(- x + x_{j+1}\right) \left(x_{j+1} - x_{j}\right)^{2} + \left(x - x_{j+1}\right)^{3}\right)}{6} + y_{j+1} \left(x - x_{j}\right) - y_{j} \left(x - x_{j+1}\right)}{x_{j+1} - x_{j}}$$

While this has been improved, there is still more which can be manually
done. For instance, the equivalent of the \(C\) term has a common factor
of \((x_{j+1} - x)\) and the \(D\) has \((x - x_j)\). \texttt{Sympy}
also tends to reorder variables, resulting in a change of signs. Some
manual work results in the following variation.

\begin{align}
y &= \frac{y_j(x_{j+1} - x) + y_{j+1}(x - x_j) + \frac 1 6 V}{x_{j+1} - x_j}, \\
V &= y''_j (x_{j+1} - x)\left((x_{j+1}-x)^2 - (x_{j+1}-x_j)^2\right) + y''_{j+1}(x-x_j)\left((x-x_j)^2 - (x_{j+1} - x_j)^2\right)
\end{align}

We can use \texttt{Sympy} to verify the original and variation are
equivalent.

\begin{lstlisting}[language=Python,xleftmargin=20pt,xrightmargin=5pt,belowskip=5pt,aboveskip=5pt]
variant_y = ( yj(0)*(xj(1) - x) + yj(1)*(x - xj(0)) + \
             (yppj(0)*(xj(1) - x)*((xj(1)-x)**2 - (xj(1)-xj(0))**2) + \
              yppj(1)*(x-xj(0))*((x-xj(0))**2 - (xj(1) - xj(0))**2))/6 ) / ( xj(1) - xj(0) )

compare_math( variant_y, simpler_y.rhs )
\end{lstlisting}

\begin{table}[H]
\centering
\begin{adjustbox}{max width=\textwidth}
The two statements are equivalent.
\end{adjustbox}
\end{table}

It does not follow that this variation is an improvement, though. The
only proper test is to convert it into computer code, and compare it
against the listing in \emph{Numerical Recipies}. For obvious reasons,
the latter code cannot be included here.

\label{code:newint}
\begin{lstlisting}[language=Python,xleftmargin=20pt,xrightmargin=5pt,belowskip=5pt,aboveskip=5pt]
# Released under a CC0 license by Haysn Hornbeck
def newint( x, a, b, u, v, up, vp ):

    assert b > a

    ba = (b - a)
    xa = (x - a)
    inv_ba = 1. / ba
    bx = (b - x)
    ba2 = ba * ba   # 3 adds, 1 mult, 1 div

    lower = xa*v + bx*u
    C = (xa*xa - ba2)*xa*vp
    D = (bx*bx - ba2)*bx*up # 1 add, 2 subs, 8 mults

    # 2 adds, 2 mult = 19 ops + 1 div
    return ( lower + (.16666666666666666666)*( C + D ) ) * inv_ba
\end{lstlisting}

In theory, we could check this algorithm by substituting \texttt{SymPy}
symbols into it, subtracting the simplified version, and cancelling
terms. In practice, \texttt{Sympy} was unable to even though a visual
inspection supports equivalence. We can still verify the algorithm via
substitution, but this requires being able to generate derivatives. We
defer on this for now.

If we ignore the initial bisection, the \emph{Numerical Recipies} code
consisted of three additions, five subtractions, ten multiplications,
and three divisions. One of those divisions was of a constant, though,
so in theory a compiler could convert it to a multiply, in which case
nineteen basic operations plus two divisions would be done.

The new code consists of six additions, two subtractions, eleven
multiplications, and one division, improving on \emph{Numerical Recipes}
by a single division. It also attempts to interleave operations so that
the result of the next calculation does not require the input of
another. This is redundant on desktop CPUs and modern compilers, which
reorder operations, but could be useful on GPUs or mobile CPUs with poor
compilers.

In practice, users of \texttt{newint} will likely translate it into C or
C++ to maximize performance. Compiler Explorer is an ideal way to
explore how a compiled language would deal with this
code.\cite{godbolt2017compiler} Here is the assembly output of
\texttt{newint} converted to C++ and compiled with \texttt{clang}
version 9.0.0, with the \texttt{-O2\ -msse4.1} optimization
flags.\footnote{-O3 gives identical output.}

\begin{mdframed}

\begin{verbatim}
.LCPI0_0:
    .long   1065353216              # float 1
.LCPI0_1:
    .long   1042983595              # float 0.166666672
newint(float, float, float, float, float, float, float):
    movaps  xmm7, xmm0
    insertps        xmm7, xmm2, 16  # xmm7 = xmm7[0],xmm2[0],xmm7[2,3]
    subss   xmm2, xmm1
    movss   xmm8, dword ptr [rip + .LCPI0_0] # xmm8 = mem[0],zero,zero,zero
    divss   xmm8, xmm2
    insertps        xmm1, xmm0, 16  # xmm1 = xmm1[0],xmm0[0],xmm1[2,3]
    subps   xmm7, xmm1
    mulss   xmm2, xmm2
    insertps        xmm4, xmm3, 16  # xmm4 = xmm4[0],xmm3[0],xmm4[2,3]
    mulps   xmm4, xmm7
    movshdup        xmm1, xmm4      # xmm1 = xmm4[1,1,3,3]
    addss   xmm1, xmm4
    movaps  xmm0, xmm7
    mulps   xmm0, xmm7
    movsldup        xmm2, xmm2      # xmm2 = xmm2[0,0,2,2]
    subps   xmm0, xmm2
    mulps   xmm0, xmm7
    insertps        xmm6, xmm5, 16  # xmm6 = xmm6[0],xmm5[0],xmm6[2,3]
    mulps   xmm6, xmm0
    movshdup        xmm0, xmm6      # xmm0 = xmm6[1,1,3,3]
    addss   xmm0, xmm6
    mulss   xmm0, dword ptr [rip + .LCPI0_1]
    addss   xmm0, xmm1
    mulss   xmm0, xmm8
    ret

\end{verbatim}

\end{mdframed}

In total there is nine instructions which move data between registers or
memory (\texttt{movaps}, \texttt{insertps}, \texttt{movshdup}), three
additions (\texttt{addss}), three subtractions (\texttt{subps},
\texttt{subss}), seven multiplications (\texttt{mulps}, \texttt{mulss}),
and one division (\texttt{divss}). It is possible to eliminate the need
to load \texttt{LCPI0\_0} into a register by removing \texttt{inv\_ba}
and simply dividing the return by \texttt{ba}, which also eliminates a
multiplication, though this moves the division to just before the return
statement and may cause execution to stall.

Compiling with \texttt{gcc} version 9.2 gives similar results, this time
with seven data moves (\texttt{movaps}, \texttt{movss}), three
additions, five subtractions (\texttt{subss}), eleven multiplies
(\texttt{mulss}), and one division. \texttt{gcc} will order the division
to occur just before \texttt{ret} with these compiler options. Some
experimentation reveals that declaring \texttt{inv\_ba} to be
\texttt{volatile} pins the division in place, at the cost adding two
extra memory moves. As the location is likely stored in L1, this could
still lead to a net speed-up as is gives the processor maximal
flexibility in scheduling the division.

The compiled results of \texttt{splint\_one}, \emph{Numerical Recipes}'
equivalent algorithm, result in one less instruction relative to the
above, though \texttt{gcc}'s output contains three divisions.
\texttt{clang} is capable of simultaneously dividing two separate
numbers by a third, unlike \texttt{gcc}, so it only requires two
divisions. As astute readers may have noticed, neither \texttt{gcc} nor
\texttt{clang} automatically converted division by a constant into a
multiplication. The authors of both compilers are aware of how
floating-point precision can change depending on the operation, and
because of that refuse to change floating-point operations unless the
\texttt{-ffast-math} flag is present. Here we know the precision loss is
trivial, so this division can easily be eliminated without needing that
flag.

\hypertarget{generating-second-derivatives}{%
\section{Generating Second
Derivatives}\label{generating-second-derivatives}}

\emph{Numerical Recipes}' interpolation algorithm assumes that each data
point's second derivative is known, something which is rarely true.
Their solution is to include \texttt{spline}, a second algorithm that
generates those second derivatives by satisfying the following equation
for all values of \(j\):

\begin{lstlisting}[language=Python,numbers=none,xleftmargin=20pt,xrightmargin=5pt,belowskip=5pt,aboveskip=5pt]
initial_value_prob = sp.Eq( (xj(0) - xj(-1))*yppj(-1) + 2*(xj(1)-xj(-1))*yppj(0) + (xj(1)-xj(0))*yppj(1), \
            6*((yj(1)-yj(0))/(xj(1)-xj(0)) - (yj(0)-yj(-1))/(xj(0)-xj(-1))) )
peq( initial_value_prob )
\end{lstlisting}

\begin{align}\label{eqn:init_val}
y''_{j+1} \left(x_{j+1} - x_{j}\right) + y''_{j-1} \left(- x_{j-1} + x_{j}\right) + y''_{j} \left(2 x_{j+1} - 2 x_{j-1}\right) = - \frac{6 \left(- y_{j-1} + y_{j}\right)}{- x_{j-1} + x_{j}} + \frac{6 \left(y_{j+1} - y_{j}\right)}{x_{j+1} - x_{j}}
\end{align}

The above equation has three unknowns, \(y''_{j+k}, k \in \{-1,0,1\}\).
As they point out, those unknowns form a tri-diagonal system. In matrix
notation, this is equivalent to

\begin{equation} \begin{bmatrix}
{b_{1}} & {c_{1}} & {} & {} & {0} \\
{a_{2}} & {b_{2}} & {c_{2}} & {} & {} \\
{} & {a_{3}} & {b_{3}} & \ddots & {} \\
{} & {} & \ddots & \ddots & {c_{n-1}} \\
{0} & {} & {} & {a_{n}} & {b_{n}} \\
\end{bmatrix} \cdot \begin{bmatrix}
{y''_{1}} \\
{y''_{2}} \\
{y''_{3}} \\
\vdots \\
{y''_{n}} \\
\end{bmatrix} ~=~ \begin{bmatrix}
{d_{1}} \\
{d_{2}} \\
{d_{3}} \\
\vdots \\
{d_{n}}\\
\end{bmatrix} \label{}\end{equation}

From the above, it is obvious that

\begin{align}
a_j &= x_j - x_{j-1} \\
b_j &= 2(x_{j+1} - x_{j-1}) \\
c_j &= x_{j+1} - x_j \\
d_j &= 6 \left( \frac{y_{j+1} - y_{j}}{x_{j+1} - x_j} - \frac{y_j - y_{j-1}}{x_j - x_{j-1}} \right)
\end{align}

for \(2 \leq j < n\). A common approach is to solve this via Thomas'
algorithm, which operates in \(O(n)\) time.\cite{thomas1949elliptic} We
must show that the above system is either diagonally dominant or
positive definite, however, in order to invoke Thomas' algorithm. The
former is easiest, as it means proving that

\begin{align}
|b_j| &\geq |a_j| + |c_j| \\
|2(x_{j+1} - x_{j-1})| &\geq |x_j - x_{j-1}| + |x_{j+1} - x_j| \label{eqn:abs_diagdom}
\end{align}

Note that the knots are written in increasing order, such that
\(x_{j+1} \geq x_j\). Thus \(|x_{j+1} - x_j| = x_{j+1} - x_j\), and we
can simplify Equation \ref{eqn:abs_diagdom} to

\begin{align}
2(x_{j+1} - x_{j-1}) &\geq x_j - x_{j-1} + x_{j+1} - x_j \label{eqn:prove_diagdom} \\
2(x_{j+1} - x_{j-1}) &\geq x_{j+1} - x_{j-1} \\
2 &\geq 1
\end{align}

As this is true for \(2 \leq j < n\), we have established diagonal
dominance in that range.

We still need values for \(b_1, c_1, d_1, a_n, b_n,\) and \(d_n\),
however. If we try to use Equation \ref{eqn:init_val} to satisfy the
first two, we find

\begin{lstlisting}[language=Python,numbers=none,xleftmargin=20pt,xrightmargin=5pt,belowskip=5pt,aboveskip=5pt]
# set up the numbered values we're substituting in
temp_x = ""
temp_y = ""
temp_ypp = ""
for v in range(0,3):
    temp_x   += 'x_{{{}}} '.format(v)
    temp_y   += 'y_{{{}}} '.format(v)
    temp_ypp += "y''_{{{}}} ".format(v)

knots_lower = sp.symbols(temp_x, real=True)
values_lower = sp.symbols(temp_y, real=True)
accels_lower = sp.symbols(temp_ypp, real=True)

del temp_x
del temp_y
del temp_ypp

def xv(index):
    return knots_lower[index]
def yv(index):
    return values_lower[index]
def yppv(index):
    return accels_lower[index]
\end{lstlisting}

\begin{lstlisting}[language=Python,numbers=none,xleftmargin=20pt,xrightmargin=5pt,belowskip=5pt,aboveskip=5pt]
# do the actual substitution
lower_prob = initial_value_prob.subs( \
    [(xj(j-1), xv(j)) for j in range(0,3)] + \
    [(yj(j-1), yv(j)) for j in range(0,3)] + \
    [(yppj(j-1), yppv(j)) for j in range(0,3)] \
        )

peq( lower_prob )
\end{lstlisting}

\begin{align}\label{eqn:lower_prob}
y''_{0} \left(- x_{0} + x_{1}\right) + y''_{1} \left(- 2 x_{0} + 2 x_{2}\right) + y''_{2} \left(- x_{1} + x_{2}\right) = \frac{6 \left(- y_{1} + y_{2}\right)}{- x_{1} + x_{2}} - \frac{6 \left(- y_{0} + y_{1}\right)}{- x_{0} + x_{1}}
\end{align}

which contains numerous undefined values. A common workaround is to
state that \(x_j = x_1\) and \(y_j = y_1\) for \(j < 1\), which reduces
the continuity at one end point. Careful substitution spares us from
both an undefined division and having to define \(y''_0\).

\begin{lstlisting}[language=Python,numbers=none,xleftmargin=20pt,xrightmargin=5pt,belowskip=5pt,aboveskip=5pt]
peq( lower_prob.subs( [(yv(0),yv(1)), (xv(0),xv(1))] ) )
\end{lstlisting}

\begin{equation}
y''_{1} \left(- 2 x_{1} + 2 x_{2}\right) + y''_{2} \left(- x_{1} + x_{2}\right) = \frac{6 \left(- y_{1} + y_{2}\right)}{- x_{1} + x_{2}}
\end{equation}

This allows us to extract the following values.

\begin{align}
b_1 &= 2(x_2 - x_1) \\
c_1 &= x_2 - x_1 \\
d_1 &= 6 \frac{y_2 - y_1}{x_2 - x_1}
\end{align}

Note that the matrix continues to be diagonally dominant, as
\(b_1 > c_1\).

\(a_n, b_n,\) and \(d_n\) are handled in a similar manner. This time we
declare \(x_j = x_n\) and \(y_j = y_n\) for \(j > n\), and again exploit
a cancellation.

\begin{lstlisting}[language=Python,numbers=none,xleftmargin=20pt,xrightmargin=5pt,belowskip=5pt,aboveskip=5pt]
# set up the offset values we're substituting in
temp_x = ""
temp_y = ""
temp_ypp = ""
for v in range(-1,2):
    base = "n"
    if v > 0:
        base += "-{}".format(v)
    elif v < 0:
        base += "+{}".format(-v)
    temp_x   += 'x_{{{}}} '.format(base)
    temp_y   += 'y_{{{}}} '.format(base)
    temp_ypp += "y''_{{{}}} ".format(base)

knots_upper = sp.symbols(temp_x, real=True)
values_upper = sp.symbols(temp_y, real=True)
accels_upper = sp.symbols(temp_ypp, real=True)

del temp_x
del temp_y
del temp_ypp

def xn(index):
    return knots_upper[1-index]
def yn(index):
    return values_upper[1-index]
def yppn(index):
    return accels_upper[1-index]
\end{lstlisting}

\begin{lstlisting}[language=Python,numbers=none,xleftmargin=20pt,xrightmargin=5pt,belowskip=5pt,aboveskip=5pt]
# do the actual substitution
upper_prob = initial_value_prob.subs( \
    [(xj(1-j), xn(1-j)) for j in range(0,3)] + \
    [(yj(1-j), yn(1-j)) for j in range(0,3)] + \
    [(yppj(1-j), yppn(1-j)) for j in range(0,3)] \
        )

peq( upper_prob.subs( [(yn(1), yn(0)), (xn(1), xn(0))] ) )
\end{lstlisting}

\begin{align}\label{eqn:upper_prob}
y''_{n-1} \left(- x_{n-1} + x_{n}\right) + y''_{n} \left(- 2 x_{n-1} + 2 x_{n}\right) = - \frac{6 \left(- y_{n-1} + y_{n}\right)}{- x_{n-1} + x_{n}}
\end{align}

Converting to the variables within the matrix, we find

\begin{align}
a_n &= x_n - x_{n-1} \\
b_n &= 2(x_n - x_{n-1}) \\
d_n &= -6 \frac{y_n - y_{n-1}}{x_n - x_{n-1}}
\end{align}

With all matrix variables filled in, we have confirmed diagonal
dominance for the entire system. Thomas' algorithm can be used to solve
it.

\hypertarget{implementation}{%
\subsection{Implementation}\label{implementation}}

That algorithm can be summarized as follows.

\begin{align}
c'_1 &= \frac{ c_1 }{ b_1 } \\
d'_1 &= \frac{ d_1 }{ b_1 } \\
c'_i &= \frac{ c_i }{ b_i - a_i c'_{i-1} } \\
d'_i &= \frac{ d_i - a_i d'_{i-1} }{ b_i - a_i c'_{i-1} } \\
y''_n &= d'_n \\
y''_i &= d'_i - c'_i y''_{i+1}
\end{align}

Note that \(a_{j+1} = c_j\), and \(d_j\) also reuses prior values in
part. As we advance \(j\), the only new values we need to fetch from
memory are \(x_{j+1}\) and \(y_{j+1}\). Everything else is either
equivalent to existing values, or can be calculated from them. While the
algorithm suggests two temporary lists, we can reduce that to one by
temporarily storing \(d'_j\) in \(y''_j\).

Some of these values can be simplified, based on the above calculations.

\begin{align}
c'_1 &= \frac{c_1}{b_1} = \frac 1 2 \\
d'_1 &= \frac{d_1}{b_1} = 3 \frac{y_2 - y_1}{(x_2 - x_1)^2}
\end{align}

We have enough information to begin writing an implementation in Python
code.

\label{code:new_second_derivative_simple}
\begin{lstlisting}[language=Python,numbers=none,xleftmargin=20pt,xrightmargin=5pt,belowskip=5pt,aboveskip=5pt]
# Released under a CC0 license by Haysn Hornbeck
def new_second_derivative_simple( knots, values ):

    assert len(knots) == len(values)
    assert len(knots) > 2

    n            = len(knots)
    c_p          = [0] * n
    ypp          = [0] * n

    # recycle these values in later routines
    new_x = knots[1]
    new_y = values[1]
    cj = knots[1] - knots[0]
    new_dj = (values[1] - values[0]) / cj

    # initialize the forward substitution
    c_p[0]  = 0.5
    ypp[0] = 3 * new_dj / cj

    # forward substitution portion
    j = 1
    while j < (n-1):

        # shuffle new values to old
        old_x = new_x
        old_y = new_y

        aj = cj
        old_dj = new_dj

        # generate new quantities
        new_x = knots[j+1]
        new_y = values[j+1]

        cj = new_x - old_x
        new_dj = (new_y - old_y) / cj
        bj = 2*(cj + aj)
        inv_denom = 1. / (bj - aj*c_p[j-1])
        dj = 6*(new_dj - old_dj)

        ypp[j] = (dj - aj*ypp[j-1]) * inv_denom
        c_p[j] = cj * inv_denom

        j += 1

    # manually do the last round, as it saves some comparisons
    old_x = new_x
    old_y = new_y

    aj = cj
    old_dj = new_dj

    cj = 0    # this has the same effect as skipping c_n and altering d_n
    new_dj = 0
    bj = 2*(cj + aj)
    inv_denom = 1. / (bj - aj*c_p[j-1])
    dj = 6*(new_dj - old_dj)

    ypp[j] = (dj - aj*ypp[j-1]) * inv_denom
    c_p[j] = cj * inv_denom

    # as we're storing d_j in y''_j, y''_n = d_n is a no-op

    # backward substitution portion
    while j > 0:

        j -= 1
        ypp[j] = ypp[j] - c_p[j]*ypp[j+1]

    return ypp
\end{lstlisting}

This is very memory-efficient, at the cost of requiring a non-trivial
number of registers. For RISC or 64-bit architectures, this usually
isn't a problem. GPU architectures may benefit from a rewrite that uses
fewer variables and more memory fetches, as registers are at a premium
while their wide memory bandwidth encourages cache use. Alternatively,
the literature on fast GPU tridiagonal solvers may yield a faster
algorithm.\cite{zhang2010fast}

While not feature-complete yet, this code is sufficiently developed to
benchmark. We have tranlated \texttt{spline} into Python for this
purpose.

\begin{lstlisting}[language=Python,numbers=none,xleftmargin=20pt,xrightmargin=5pt,belowskip=5pt,aboveskip=5pt]
import numpy as np
import timeit

test_knots  = np.linspace( 0, np.pi*2, 256 )
test_values = np.random.rand( 256 )

bench = timeit.Timer( lambda: spline(test_knots, test_values, 1e31, 1e31) )
count, time = bench.autorange()
count *= 20  # inflate the runtime from about 0.2s to about 4s

result_nr = bench.timeit(number=count)

print("Time to generate second derivatives via NR's code:\t{:.3f}s ({} trials)".format( result_nr, count ))

bench = timeit.Timer( lambda: new_second_derivative_simple( test_knots, test_values ) )
result_sk = bench.timeit(number=count)

print("Time to generate second derivatives via the sketch:\t{:.3f}s ({} trials)".format( result_sk, count ))

print("  Speedup: approx {:.2f}x".format(result_nr / result_sk) )
\end{lstlisting}

\begin{lstlisting}[language={},postbreak={},numbers=none,xrightmargin=7pt,belowskip=5pt,aboveskip=5pt,breakindent=0pt]
Time to generate second derivatives via NR's code:	4.899s (4000 trials)
Time to generate second derivatives via the sketch:	2.809s (4000 trials)
  Speedup: approx 1.74x

\end{lstlisting}

The new algorithm is faster by a non-trivial amount. It's possible that
switching to a compiled language could narrow the difference, by
reducing memory fetches and optimizing the underlying code. Since
generating second derivatives is unlikely to be a performance bottleneck
in practice, this line of investigation is not worth pursuing.

\hypertarget{start-and-end-derivatives}{%
\subsection{Start and End Derivatives}\label{start-and-end-derivatives}}

\texttt{new\_second\_derivative\_simple} is not a drop-in replacement
for \emph{Numerical Recipes}' \texttt{spline}, unfortunately. The latter
also provides a way to alter the slope at the start of the curve, the
end, or both places.

Earlier we declared that \(x_j = x_1\) and \(y_j = y_1\) for \(j < 1\),
in order to cancel out a term in the matrix representation of this
problem. Another approach is to declare \(y''_j = 0\) for \(j < 1\).

\begin{lstlisting}[language=Python,numbers=none,xleftmargin=20pt,xrightmargin=5pt,belowskip=5pt,aboveskip=5pt]
lower_prob_sub = lower_prob.subs( yppv(0),0 )

peq( lower_prob_sub )
\end{lstlisting}

\begin{align}\label{eqn:lower_deriv_base}
y''_{1} \left(- 2 x_{0} + 2 x_{2}\right) + y''_{2} \left(- x_{1} + x_{2}\right) = \frac{6 \left(- y_{1} + y_{2}\right)}{- x_{1} + x_{2}} - \frac{6 \left(- y_{0} + y_{1}\right)}{- x_{0} + x_{1}}
\end{align}

While this may not seem like progress, observe that
\(\frac{y_1 - y_0}{x_1 - x_0}\) is a good approximation of the slope
between two knots on the curve. The code in \emph{Numerical Recipes} in
fact simply replaces that term with the desired starting derivative,
which we will denote as \(y'_1\), and asserts \(x_0 = x_1\). It is worth
re-deriving this result, if only to check it results in the correct
derivative.

We begin with evaluating Equation \ref{eqn:nr_formula} at \(j = 0\),
taking its derivative at \(x = x_1\), and asserting \(y''_j = 0\) for
\(j < 1\).

\begin{lstlisting}[language=Python,numbers=none,xleftmargin=20pt,xrightmargin=5pt,belowskip=5pt,aboveskip=5pt]
yp1 = sp.Symbol( "y'_1", real=True )

variant_y_first = variant_y.subs( \
    [(xj(j), xv(j)) for j in range(0,2)] + \
    [(yj(j), yv(j)) for j in range(0,2)] + \
    [(yppj(j), yppv(j)) for j in range(0,2)] \
        )

var_y_first_deriv = sp.Eq( yp1, sp.simplify( sp.diff( variant_y_first, x ).subs( [(yppv(0), 0),(x,xv(1))] ) ) )

peq( var_y_first_deriv )
\end{lstlisting}

\begin{equation}
y'_{1} = \frac{- \frac{y''_{1} \left(x_{0} - x_{1}\right)^{2}}{3} + y_{0} - y_{1}}{x_{0} - x_{1}}
\end{equation}

Unfortunately, this is an equation with three unknowns. If we assume
that \(x_2 > x_1\), however, we can assert that
\(x_0 = x_1 - (x_2 - x_1) = 2x_1 - x_2\). This will allow us to solve
for \(y_0\).

\begin{lstlisting}[language=Python,numbers=none,xleftmargin=20pt,xrightmargin=5pt,belowskip=5pt,aboveskip=5pt]
var_y_first_subs = var_y_first_deriv.subs( xv(0), 2*xv(1) - xv(2) )
var_y_first_sol = sp.solve( var_y_first_subs, yv(0) )[0].collect( yp1 )

sp.Eq( yv(0), var_y_first_sol )
\end{lstlisting}

\begin{align}\label{eqn:lower_deriv_sub}
\displaystyle y_{0} = \frac{y''_{1} \left(x_{1} - x_{2}\right)^{2}}{3} + y'_{1} \left(x_{1} - x_{2}\right) + y_{1}
\end{align}

We can now substitute Equation \ref{eqn:lower_deriv_sub} into Equation
\ref{eqn:lower_deriv_base}, along with our other assertions.

\begin{lstlisting}[language=Python,numbers=none,xleftmargin=20pt,xrightmargin=5pt,belowskip=5pt,aboveskip=5pt]
lower_prob_int = sp.simplify( lower_prob_sub.subs( [(xv(0), 2*xv(1) - xv(2)), (yv(0), var_y_first_sol)] ) )
lower_prob_tidy = sp.Eq( yppv(1), sp.solve( lower_prob_int, yppv(1) )[0].collect( yp1 ) )
peq( lower_prob_tidy )
\end{lstlisting}

\begin{align}\label{eqn:lower_deriv_machine}
y''_{1} = \frac{- y''_{2} \left(x_{1} - x_{2}\right)^{2} + y'_{1} \left(6 x_{1} - 6 x_{2}\right) - 6 y_{1} + 6 y_{2}}{2 \left(x_{1} - x_{2}\right)^{2}}
\end{align}

\texttt{Sympy} has difficulty rearranging this equation to match
Equation \ref{eqn:init_val}, but the task is trivial for a human.

\begin{lstlisting}[language=Python,numbers=none,xleftmargin=20pt,xrightmargin=5pt,belowskip=5pt,aboveskip=5pt]
lower_prob_sol = sp.Eq( 2*yppv(1)*(xv(2) - xv(1)) + yppv(2)*(xv(2) - xv(1)), \
                        6*((yv(2) - yv(1))/(xv(2) - xv(1)) - yp1) )

peq( lower_prob_sol )
compare_math( sp.solve(lower_prob_sol, yppv(1))[0], lower_prob_tidy.rhs )
\end{lstlisting}

\begin{equation}
2 y''_{1} \left(- x_{1} + x_{2}\right) + y''_{2} \left(- x_{1} + x_{2}\right) = - 6 y'_{1} + \frac{6 \left(- y_{1} + y_{2}\right)}{- x_{1} + x_{2}}
\end{equation}

\begin{table}[H]
\centering
\begin{adjustbox}{max width=\textwidth}
The two statements are equivalent.
\end{adjustbox}
\end{table}

Converting this to matrix form, we find

\begin{align}
c'_1 &= \frac{c_1}{b_1} = \frac{x_2 - x_1}{2(x_2 - x_1)} = \frac 1 2 \\
d'_1 &= \frac{d_1}{b_1} = \frac{6 \left(\frac{y_2 - y_1}{x_2 - x_1} - y'_1\right)}{2(x_2 - x_1)} = \frac 3 {x_2 - x_1} \left( \frac{y_2 - y_1}{x_2 - x_1} - y'_1 \right)
\end{align}

which matches \emph{Numerical Recipes} exactly. Similar logic applies to
the solution at the other end, this time we assert \(y''_{j} = 0\) for
\(j > n\) and \(x_{n+1} = x_n + (x_n - x_{n-1}) = 2x_n - x_{n-1}\).

\begin{lstlisting}[language=Python,numbers=none,xleftmargin=20pt,xrightmargin=5pt,belowskip=5pt,aboveskip=5pt]
ypn = sp.Symbol( "y'_n", real=True )

upper_prob_sub = upper_prob.subs( yppn(1),0 )

# convert the generic case to a specific one
variant_y_last = variant_y.subs( \
    [(xj(j), xn(j)) for j in range(0,2)] + \
    [(yj(j), yn(j)) for j in range(0,2)] + \
    [(yppj(j), yppn(j)) for j in range(0,2)] \
        )

# differentiate and substitute into the specific case
var_y_last_deriv = sp.Eq( ypn, sp.simplify( sp.diff( variant_y_last, x ).subs( [(yppn(1), 0),(x,xn(0))] ) ) )

# perform more substitutions and simplify
var_y_last_subs = var_y_last_deriv.subs( xn(1), 2*xn(0) - xn(-1) )
var_y_last_sol = sp.solve( var_y_last_subs, yn(1) )[0].collect( ypn )

# substitute one equation into the other, redo some earlier substitutions, and tidy it up
upper_prob_int = sp.simplify( upper_prob_sub.subs( [(xn(1), 2*xn(0) - xn(-1)), (yn(1), var_y_last_sol)] ) )
upper_prob_tidy = sp.Eq( yppn(0), sp.solve( upper_prob_int, yppn(0) )[0].collect( ypn ) )

peq( upper_prob_tidy )
\end{lstlisting}

\begin{equation}
y''_{n} = \frac{- y''_{n-1} \left(x_{n-1} - x_{n}\right)^{2} + y'_{n} \left(- 6 x_{n-1} + 6 x_{n}\right) + 6 y_{n-1} - 6 y_{n}}{2 \left(x_{n-1} - x_{n}\right)^{2}}
\end{equation}

Again, some manual tidying is necessary.

\begin{lstlisting}[language=Python,numbers=none,xleftmargin=20pt,xrightmargin=5pt,belowskip=5pt,aboveskip=5pt]
upper_prob_sol = sp.Eq( 2*yppn(0)*(xn(0) - xn(-1)) + yppn(-1)*(xn(0) - xn(-1)), \
                        6*(ypn - (yn(0) - yn(-1))/(xn(0) - xn(-1))) )

peq( upper_prob_sol )
compare_math( sp.solve(upper_prob_sol, yppn(0))[0], upper_prob_tidy.rhs )
\end{lstlisting}

\begin{equation}
y''_{n-1} \left(- x_{n-1} + x_{n}\right) + 2 y''_{n} \left(- x_{n-1} + x_{n}\right) = 6 y'_{n} - \frac{6 \left(- y_{n-1} + y_{n}\right)}{- x_{n-1} + x_{n}}
\end{equation}

\begin{table}[H]
\centering
\begin{adjustbox}{max width=\textwidth}
The two statements are equivalent.
\end{adjustbox}
\end{table}

When converted to matrix form, we again find agreement with
\emph{Numerical Recipes}, though it should be noted their code takes a
different approach to the end of the matrix than ours.

\begin{align}
a_n &= 2(x_n - x_{n-1}) \\
b_n &= x_n - x_{n-1} \\
d_n &= 6 \left( y'_n - \frac{y_n - y_{n-1}}{x_n - x_{n-1}} \right)
\end{align}

We are now capable of providing a feature-complete drop-in replacement
for \texttt{spline}.

\label{code:new_second_derivative}
\begin{lstlisting}[language=Python,numbers=none,xleftmargin=20pt,xrightmargin=5pt,belowskip=5pt,aboveskip=5pt]
# Released under a CC0 license by Haysn Hornbeck
def new_second_derivative( knots, values, start_deriv, end_deriv ):

    assert len(knots) == len(values)
    assert len(knots) > 2
    for i,_ in enumerate(knots):
        assert (i == 0) or (knots[i] > knots[i-1])

    n            = len(knots)
    c_p          = [0] * n
    ypp          = [0] * n

    # recycle these values in later routines
    new_x = knots[1]
    new_y = values[1]
    cj = knots[1] - knots[0]
    new_dj = (values[1] - values[0]) / cj

    # initialize the forward substitution
    if start_deriv > .99e30:
        c_p[0] = 0
        ypp[0] = 0
    else:
        c_p[0]  = 0.5
        ypp[0] = 3 * (new_dj - start_deriv) / cj

    # forward substitution portion
    j = 1
    while j < (n-1):

        # shuffle new values to old
        old_x = new_x
        old_y = new_y

        aj = cj
        old_dj = new_dj

        # generate new quantities
        new_x = knots[j+1]
        new_y = values[j+1]

        cj = new_x - old_x
        new_dj = (new_y - old_y) / cj
        bj = 2*(cj + aj)
        inv_denom = 1. / (bj - aj*c_p[j-1])
        dj = 6*(new_dj - old_dj)

        ypp[j] = (dj - aj*ypp[j-1]) * inv_denom
        c_p[j] = cj * inv_denom

        j += 1

    # handle the end derivative
    if end_deriv > .99e30:
        c_p[j] = 0
        ypp[j] = 0

    else:
        old_x = new_x
        old_y = new_y

        aj = cj
        old_dj = new_dj

        cj = 0    # this has the same effect as skipping c_n
        new_dj = end_deriv
        bj = 2*(cj + aj)
        inv_denom = 1. / (bj - aj*c_p[j-1])
        dj = 6*(new_dj - old_dj)

        ypp[j] = (dj - aj*ypp[j-1]) * inv_denom
        c_p[j] = cj * inv_denom

    # as we're storing d_j in y''_j, y''_n = d_n is a no-op

    # backward substitution portion
    while j > 0:

        j -= 1
        ypp[j] = ypp[j] - c_p[j]*ypp[j+1]

    return ypp
\end{lstlisting}

As the inner loop of this code is no different from
\texttt{new\_second\_derivative\_simple}, it will offer similar
performance.

\hypertarget{quality-checks}{%
\section{Quality Checks}\label{quality-checks}}

\begin{figure*}[tbh]\begin{center}\adjustimage{max size={0.9\linewidth}{0.9\paperheight}}{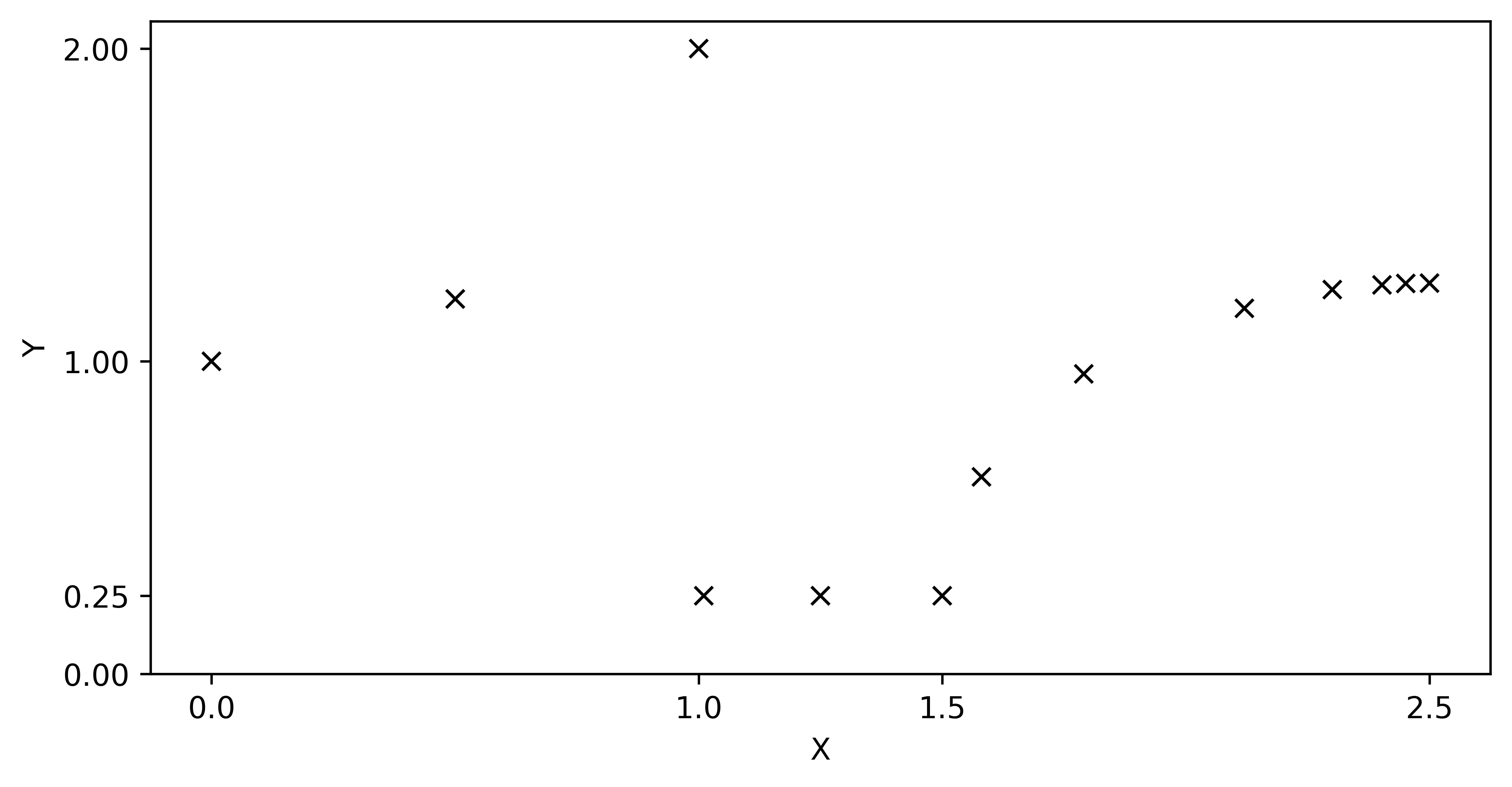}\end{center}\caption{The hand-generated control points.}\label{fig:hand_points}\end{figure*}

Since our goal is to create a drop-in replacement, we must be assured
the new code generates the same results as the old. One approach is to
hand-generate a set of control points and observe what happens when both
routines are applied to it. The curve we will use is designed to have a
mix of close and distant knots, sharp edges and smooth transitions.
Figure \ref{fig:hand_points} plots its control points.

\begin{lstlisting}[language=Python,numbers=none,xleftmargin=20pt,xrightmargin=5pt,belowskip=5pt,aboveskip=5pt]
test_knots  = [0,  0.5, 1, 1.01, 1.25,   1.5, 1.58, 1.79, 2.12, 2.30, 2.402, 2.451, 2.5]
test_values = [1,  1.2, 2, 0.25, 0.25,  0.25, 0.63, 0.96, 1.17, 1.23, 1.245, 1.249, 1.25]

%matplotlib inline
import matplotlib.pyplot as plt
plt.figure(num=None, figsize=(8, 4), dpi=600, facecolor='w', edgecolor='k')

plt.plot( test_knots, test_values, 'xk' )
plt.xlabel( "X" )
plt.ylabel( "Y" )
plt.xticks( [0,1,1.5,2.5] )
plt.yticks( [0,0.25,1,2] )
plt.show()
\end{lstlisting}

The simplest quantitative comparison is to calculate the second
derivatives with both \texttt{spline} and
\texttt{new\_second\_derivative}, and create a table of the different
values calculated by each routine. There are three important cases to
test: ``natural'' cubic b-splines that have \(y''_1 = y''_n = 0\), cubic
b-splines with \(y'_1 = y'_n = 0\), and cubic b-splines with non-trivial
derivatives. For the latter case, we will use \(y'_1 = -1\) to
exaggerate the curvature at the beginning of the curve and \(y'_n = 1\)
to skew what would otherwise be a nearly flat portion of the curve.

We begin by comparing what each routine generates when asked for a
``natural'' cubic b-spline.

\begin{lstlisting}[language=Python,numbers=none,xleftmargin=20pt,xrightmargin=5pt,belowskip=5pt,aboveskip=5pt]
test_secdir_old = spline( test_knots, test_values, 1e32, 1e32 )
test_secdir_new = new_second_derivative( test_knots, test_values, 1e32, 1e32 )

import pandas as pd
table = pd.DataFrame( {'$x$':test_knots, '$y$':test_values, \
                       'NR':test_secdir_old, 'our':test_secdir_new, \
                       '$\Delta$':[a-b for a,b in zip(test_secdir_old, test_secdir_new)] } )

display( {"text/latex":table.to_latex(index=False), "text/html":table.to_html(index=False)}, raw=True )
\end{lstlisting}

\begin{table}[H]
\caption{A comparison of the two routines for a ``natural'' cubic b-spline. The
calculations were done with double-precision floating-point numbers.
Deltas are calculated by subtracting the old implementation from the
new.}\label{tbl:test_compare_nat1}
\centering
\begin{adjustbox}{max width=\textwidth}\rowcolors{2}{gray!20}{white}
\begin{tabular}{rrrrr}
\toprule
   $x$ &    $y$ &           NR &          our &      $\Delta$ \\
\midrule
 0.000 &  1.000 &     0.000000 &     0.000000 &  0.000000e+00 \\
 0.500 &  1.200 &   306.924794 &   306.924794 &  5.684342e-14 \\
 1.000 &  2.000 & -1213.299177 & -1213.299177 & -2.273737e-13 \\
 1.010 &  0.250 &  2450.276370 &  2450.276370 & -4.547474e-13 \\
 1.250 &  0.250 &  -679.188305 &  -679.188305 &  3.410605e-13 \\
 1.500 &  0.250 &   310.152841 &   310.152841 & -1.136868e-13 \\
 1.580 &  0.630 &   -80.047486 &   -80.047486 &  1.421085e-14 \\
 1.790 &  0.960 &    12.113742 &    12.113742 & -3.552714e-15 \\
 2.120 &  1.170 &    -5.706845 &    -5.706845 &  0.000000e+00 \\
 2.300 &  1.230 &     0.029250 &     0.029250 & -2.220446e-16 \\
 2.402 &  1.245 &    -1.048158 &    -1.048158 &  2.220446e-16 \\
 2.451 &  1.249 &    -1.612180 &    -1.612180 & -2.220446e-16 \\
 2.500 &  1.250 &     0.000000 &     0.000000 &  0.000000e+00 \\
\bottomrule
\end{tabular}

\end{adjustbox}
\end{table}

Table \ref{tbl:test_compare_nat1} demonstrates close agreement between
\emph{Numerical Recipes} routine and ours, with the differences at the
precision limit for double-precision floating-point numbers. Which
routine is the closest to the true value, however, if we could calculate
values with infinite precision? We can approximate an answer with the
arbitrary precision math routine \texttt{mpmath}.\cite{mpmath} The
Python versions of both \emph{Numerical Recipes}' code and ours work
transparently with this library, so the only code changes involve
generating the input arrays and adding the proper comparisons.

\begin{lstlisting}[language=Python,numbers=none,xleftmargin=20pt,xrightmargin=5pt,belowskip=5pt,aboveskip=5pt]
from mpmath import mp

test_knots_hp  = [mp.mpf('0'), mp.mpf('0.5'), mp.mpf('1'), mp.mpf('1.01'), mp.mpf('1.25'), mp.mpf('1.5'), \
                  mp.mpf('1.58'), mp.mpf('1.79'), mp.mpf('2.12'), mp.mpf('2.30'), mp.mpf('2.402'), \
                  mp.mpf('2.451'), mp.mpf('2.5')]
test_values_hp = [mp.mpf('1'), mp.mpf('1.2'), mp.mpf('2'), mp.mpf('0.25'), mp.mpf('0.25'), mp.mpf('0.25'), \
                  mp.mpf('0.63'), mp.mpf('0.96'), mp.mpf('1.17'), mp.mpf('1.23'), mp.mpf('1.245'), \
                  mp.mpf('1.249'), mp.mpf('1.25')]

precis = dict()
for precision in range(10,150,20):
    with mp.workdps(precision):
        precis[precision] = mp.nstr( np.max([(a-b) for a,b in zip( \
                                spline( test_knots_hp, test_values_hp, 1e32, 1e32 ), \
                                new_second_derivative( test_knots_hp, test_values_hp, 1e32, 1e32 ) )]), 5 )

table = pd.DataFrame( precis, index=["maximal disagreement:"] )
table.columns.name = "decimal precision:"

display( {"text/latex":table.to_latex(index=True), "text/html":table.to_html(index=True)}, raw=True )
\end{lstlisting}

\begin{table}[H]
\caption{The maximal divergence between NR code and ours, for varying degrees of
precision.}\label{tbl:arb_precision}
\centering
\begin{adjustbox}{max width=\textwidth}\rowcolors{2}{gray!20}{white}
\begin{tabular}{llllllll}
\toprule
decimal precision: &        10  &         30  &         50  &        70  &         90  &          110 &          130 \\
\midrule
maximal disagreement: &  7.4506e-9 &  2.0195e-28 &  1.3685e-48 &  9.273e-69 &  2.5135e-88 &  3.4064e-108 &  2.3082e-128 \\
\bottomrule
\end{tabular}

\end{adjustbox}
\end{table}

Table \ref{tbl:arb_precision} suggests that the precision of both
\emph{Numerical Recipes}' code and ours are dependent only on the
precision of the underlying floating-point storage. This means that we
can approximate the true output of either routine by using
\texttt{mpmath} with a precision sufficiently smaller than IEEE
binary64's typical precision of 15-17 decimal places. We have
arbitrarily chosen to use \texttt{spline} and a precision of 30 decimal
places to calculate the ``true'' value.

\begin{lstlisting}[language=Python,numbers=none,xleftmargin=20pt,xrightmargin=5pt,belowskip=5pt,aboveskip=5pt]
mp.dps = 30
test_secdir_true = spline( test_knots_hp, test_values_hp, 1e32, 1e32 )

table = pd.DataFrame( {'$x$':test_knots, '$y$':test_values, \
                       'NR $\Delta$, "true"': \
                            [mp.nstr(a-b,5) for a,b in zip(test_secdir_old, test_secdir_true)], \
                       'our $\Delta$, "true"': \
                            [mp.nstr(a-b,5) for a,b in zip(test_secdir_new, test_secdir_true)]} )

display( {"text/latex":table.to_latex(index=False), "text/html":table.to_html(index=False)}, raw=True )
\end{lstlisting}

\begin{table}[H]
\caption{A comparison of the two routines for a ``natural'' cubic b-spline. The
calculations were done with double-precision floating-point numbers,
this time comparing each to the ``true'' value.}\label{tbl:test_compare_nat}
\centering
\begin{adjustbox}{max width=\textwidth}\rowcolors{2}{gray!20}{white}
\begin{tabular}{rrll}
\toprule
   $x$ &    $y$ & NR $\Delta$, "true" & our $\Delta$, "true" \\
\midrule
 0.000 &  1.000 &                 0.0 &                  0.0 \\
 0.500 &  1.200 &          4.5507e-14 &          -1.1336e-14 \\
 1.000 &  2.000 &         -9.3212e-14 &           1.3416e-13 \\
 1.010 &  0.250 &          2.2244e-13 &           6.7719e-13 \\
 1.250 &  0.250 &          7.1667e-14 &          -2.6939e-13 \\
 1.500 &  0.250 &         -1.4856e-14 &           9.8831e-14 \\
 1.580 &  0.630 &         -5.6305e-15 &          -1.9841e-14 \\
 1.790 &  0.960 &           1.985e-15 &           5.5377e-15 \\
 2.120 &  1.170 &         -8.3121e-16 &          -8.3121e-16 \\
 2.300 &  1.230 &          5.7138e-16 &           7.9343e-16 \\
 2.402 &  1.245 &         -3.2598e-16 &          -5.4802e-16 \\
 2.451 &  1.249 &         -8.7028e-17 &           1.3502e-16 \\
 2.500 &  1.250 &                 0.0 &                  0.0 \\
\bottomrule
\end{tabular}

\end{adjustbox}
\end{table}

Table \ref{tbl:test_compare_nat} shows both the old and new routines are
remarkably close to the true value. We can quantify their overall
difference by calculating the mean square error.

\begin{lstlisting}[language=Python,numbers=none,xleftmargin=20pt,xrightmargin=5pt,belowskip=5pt,aboveskip=5pt]
table = pd.DataFrame( {'MSE, NR' :[mp.nstr(np.mean( [(a-b)**2 for a,b in zip(test_secdir_old, test_secdir_true)] ))], \
                       'MSE, our':[mp.nstr(np.mean( [(a-b)**2 for a,b in zip(test_secdir_new, test_secdir_true)] ))] })

display( {"text/latex":table.to_latex(index=False), "text/html":table.to_html(index=False)}, raw=True )
\end{lstlisting}

\begin{table}[H]
\caption{The mean square error of each set of routines from the ``true'' value,
for a ``natural'' cubic B-spline.}\label{tbl:mse_nat}
\centering
\begin{adjustbox}{max width=\textwidth}\rowcolors{2}{gray!20}{white}
\begin{tabular}{ll}
\toprule
     MSE, NR &     MSE, our \\
\midrule
 5.04867e-27 &  4.30368e-26 \\
\bottomrule
\end{tabular}

\end{adjustbox}
\end{table}

According to Table \ref{tbl:mse_nat}, \emph{Numerical Recipes}' routine
is closer to the ``true'' value than ours. It is worth noting that this
difference appears chaotic. This paper was created using an interactive
Jupyter notebook, and the minor differences between the interactive
environment and the non-interactive one used to execute the final paper
were enough to cause our routine to appear more accurate in the latter.
This is consistent with both routines being unbiased in the limit, while
for finite precision both have a bounded yet chaotic imprecision.

Next, we will compare the two routines when the initial and final slope
is a horizontal line.

\begin{lstlisting}[language=Python,numbers=none,xleftmargin=20pt,xrightmargin=5pt,belowskip=5pt,aboveskip=5pt]
test_secdir_old = spline( test_knots, test_values, 0, 0 )
test_secdir_new = new_second_derivative( test_knots, test_values, 0, 0 )
test_secdir_true = spline( test_knots_hp, test_values_hp, 0, 0 )

table = pd.DataFrame( {'$x$':test_knots, '$y$':test_values, \
                       'NR $\Delta$, "true"': \
                            [mp.nstr(a-b,5) for a,b in zip(test_secdir_old, test_secdir_true)], \
                       'our $\Delta$, "true"': \
                            [mp.nstr(a-b,5) for a,b in zip(test_secdir_new, test_secdir_true)]} )

display( {"text/latex":table.to_latex(index=False), "text/html":table.to_html(index=False)}, raw=True )
\end{lstlisting}

\begin{table}[H]
\caption{A comparison of the two routines for a cubic b-spline with
\(y'_1 = y'_n = 0\).}\label{tbl:test_compare_flat}
\centering
\begin{adjustbox}{max width=\textwidth}\rowcolors{2}{gray!20}{white}
\begin{tabular}{rrll}
\toprule
   $x$ &    $y$ & NR $\Delta$, "true" & our $\Delta$, "true" \\
\midrule
 0.000 &  1.000 &          1.7473e-14 &           1.7473e-14 \\
 0.500 &  1.200 &         -2.2512e-14 &          -2.2512e-14 \\
 1.000 &  2.000 &          4.7705e-14 &           4.7705e-14 \\
 1.010 &  0.250 &          6.9425e-15 &           6.9425e-15 \\
 1.250 &  0.250 &         -8.3012e-14 &           3.0675e-14 \\
 1.500 &  0.250 &          7.3029e-14 &           1.6186e-14 \\
 1.580 &  0.630 &         -5.6357e-15 &          -5.6357e-15 \\
 1.790 &  0.960 &          2.5403e-15 &           2.5403e-15 \\
 2.120 &  1.170 &         -9.4845e-16 &           -6.027e-17 \\
 2.300 &  1.230 &          3.9991e-16 &           1.2236e-16 \\
 2.402 &  1.245 &         -1.8866e-16 &          -1.8866e-16 \\
 2.451 &  1.249 &         -1.9206e-16 &           2.9989e-17 \\
 2.500 &  1.250 &          6.0745e-17 &           -1.613e-16 \\
\bottomrule
\end{tabular}

\end{adjustbox}
\end{table}

\begin{lstlisting}[language=Python,numbers=none,xleftmargin=20pt,xrightmargin=5pt,belowskip=5pt,aboveskip=5pt]
table = pd.DataFrame( {'MSE, NR' :[mp.nstr(np.mean( [(a-b)**2 for a,b in zip(test_secdir_old, test_secdir_true)] ))], \
                       'MSE, our':[mp.nstr(np.mean( [(a-b)**2 for a,b in zip(test_secdir_new, test_secdir_true)] ))] })

display( {"text/latex":table.to_latex(index=False), "text/html":table.to_html(index=False)}, raw=True )
\end{lstlisting}

\begin{table}[H]
\caption{The mean square error of each set of routines from the ``true'' value.}\label{tbl:mse_flat}
\centering
\begin{adjustbox}{max width=\textwidth}\rowcolors{2}{gray!20}{white}
\begin{tabular}{ll}
\toprule
     MSE, NR &     MSE, our \\
\midrule
 1.18459e-27 &  3.36714e-28 \\
\bottomrule
\end{tabular}

\end{adjustbox}
\end{table}

As Table \ref{tbl:test_compare_flat} demonstrates, these routines again
agree to floating-point precision. Table \ref{tbl:mse_flat} shows that
our routine is closer the ``true'' value than \emph{Numerical Recipes}.
The Jupyter notebook version again says the opposite.

Finally, we change the start and end derivatives to be non-trivial,
\(y'_1 = -1\) and \(y'_n = 1\).

\begin{lstlisting}[language=Python,numbers=none,xleftmargin=20pt,xrightmargin=5pt,belowskip=5pt,aboveskip=5pt]
test_secdir_old = spline( test_knots, test_values, -1, 1 )
test_secdir_new = new_second_derivative( test_knots, test_values, -1, 1 )
test_secdir_true = spline( test_knots_hp, test_values_hp, -1, 1 )

table = pd.DataFrame( {'$x$':test_knots, '$y$':test_values, \
                       'NR $\Delta$, "true"': \
                            [mp.nstr(a-b,5) for a,b in zip(test_secdir_old, test_secdir_true)], \
                       'our $\Delta$, "true"': \
                            [mp.nstr(a-b,5) for a,b in zip(test_secdir_new, test_secdir_true)]} )

display( {"text/latex":table.to_latex(index=False), "text/html":table.to_html(index=False)}, raw=True )
\end{lstlisting}

\begin{table}[H]
\caption{A comparison of the two routines for a cubic b-spline with \(y'_1 = -1\)
and \(y'_n = 1\). The calculations were done with double-precision
floating-point numbers. Deltas are calculated by subtracting the old
implementation from the new.}\label{tbl:test_compare_twist}
\centering
\begin{adjustbox}{max width=\textwidth}\rowcolors{2}{gray!20}{white}
\begin{tabular}{rrll}
\toprule
   $x$ &    $y$ & NR $\Delta$, "true" & our $\Delta$, "true" \\
\midrule
 0.000 &  1.000 &          1.4693e-14 &           4.3115e-14 \\
 0.500 &  1.200 &         -1.6952e-14 &          -7.3795e-14 \\
 1.000 &  2.000 &         -2.8599e-14 &           1.9877e-13 \\
 1.010 &  0.250 &         -4.4446e-13 &          -4.4446e-13 \\
 1.250 &  0.250 &          1.3114e-13 &           2.4483e-13 \\
 1.500 &  0.250 &          -5.115e-14 &           -5.115e-14 \\
 1.580 &  0.630 &         -5.6418e-15 &          -5.6418e-15 \\
 1.790 &  0.960 &         -1.7248e-15 &           1.8279e-15 \\
 2.120 &  1.170 &          1.6451e-15 &          -1.0194e-15 \\
 2.300 &  1.230 &         -2.9414e-16 &           3.7199e-16 \\
 2.402 &  1.245 &         -2.4384e-16 &          -6.8792e-16 \\
 2.451 &  1.249 &          1.8995e-15 &           1.8995e-15 \\
 2.500 &  1.250 &          1.7419e-15 &           1.7419e-15 \\
\bottomrule
\end{tabular}

\end{adjustbox}
\end{table}

\begin{lstlisting}[language=Python,numbers=none,xleftmargin=20pt,xrightmargin=5pt,belowskip=5pt,aboveskip=5pt]
table = pd.DataFrame( {'MSE, NR' :[mp.nstr(np.mean( [(a-b)**2 for a,b in zip(test_secdir_old, test_secdir_true)] ))], \
                       'MSE, our':[mp.nstr(np.mean( [(a-b)**2 for a,b in zip(test_secdir_new, test_secdir_true)] ))] })

display( {"text/latex":table.to_latex(index=False), "text/html":table.to_html(index=False)}, raw=True )
\end{lstlisting}

\begin{table}[H]
\caption{The mean square error of each set of routines from the ``true'' value.}\label{tbl:mse_twist}
\centering
\begin{adjustbox}{max width=\textwidth}\rowcolors{2}{gray!20}{white}
\begin{tabular}{ll}
\toprule
     MSE, NR &     MSE, our \\
\midrule
 1.68251e-26 &  2.36125e-26 \\
\bottomrule
\end{tabular}

\end{adjustbox}
\end{table}

As expected, Table \ref{tbl:test_compare_twist} shows no significant
differences between the two routines. Table \ref{tbl:mse_twist} shows
that the \emph{Numerical Recipes} code is closer to the true value, and
the interactive version agrees with this.

We will use this last set of second derivatives to perform a qualitative
check on an interpolated curve, relying on both \emph{Numerical
Recipes}' \texttt{splint\_one} and our \texttt{newint}.

\begin{lstlisting}[language=Python,numbers=none,xleftmargin=20pt,xrightmargin=5pt,belowskip=5pt,aboveskip=5pt]
x_values = np.linspace(min(test_knots), max(test_knots), 2048)
y_values_old = list()
y_values_new = list()
y_values_true = list()

i = 0
for x in x_values:
    # find the proper lower index
    while (i+1 < len(test_knots)) and (test_knots[i+1] < x):
        i += 1

    y_values_old.append( splint_one( x, test_knots[i], test_knots[i+1], \
                                    test_values[i], test_values[i+1], \
                                    test_secdir_old[i], test_secdir_old[i+1] ) )

    y_values_new.append( newint( x, test_knots[i], test_knots[i+1], \
                                         test_values[i], test_values[i+1], \
                                         test_secdir_new[i], test_secdir_new[i+1] ) )

    y_values_true.append( splint_one( x, test_knots_hp[i], test_knots_hp[i+1], \
                                    test_values_hp[i], test_values_hp[i+1], \
                                    test_secdir_true[i], test_secdir_true[i+1] ) )
\end{lstlisting}

\begin{lstlisting}[language=Python,numbers=none,xleftmargin=20pt,xrightmargin=5pt,belowskip=5pt,aboveskip=5pt]
plt.figure(num=None, figsize=(8, 4), dpi=600, facecolor='w', edgecolor='k')
plt.plot( test_knots, test_values, 'xk' )

plt.plot( x_values, y_values_old, '-', color='#0000FF7F' )
plt.plot( x_values, y_values_new, '-', color='#FF00007F' )

plt.legend( ["control points","old algorithms","new algorithms"] )
plt.xlabel("X")
plt.ylabel("Y")
plt.ylim( [-10,20] )
plt.show()
\end{lstlisting}

\begin{figure*}[tbh]\begin{center}\adjustimage{max size={0.9\linewidth}{0.9\paperheight}}{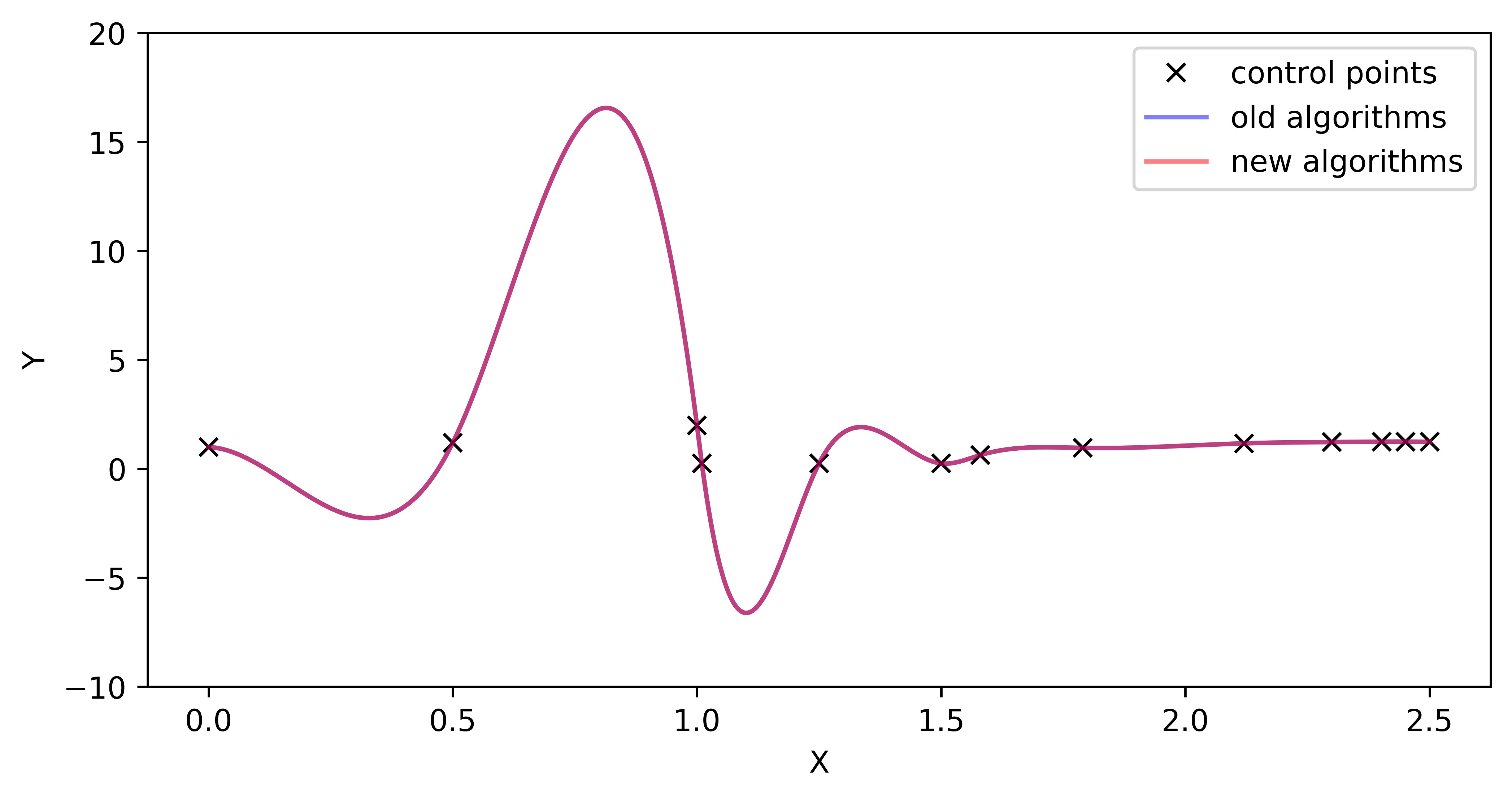}\end{center}\caption{A qualitative check of the old and new B-spline routines.}\label{fig:qual_check}\end{figure*}

As Figure \ref{fig:qual_check} shows, both sets of algorithms are in
broad agreement. The resolution is very coarse, however, and does not
answer which is closer to the ``true'' value.

\begin{lstlisting}[language=Python,numbers=none,xleftmargin=20pt,xrightmargin=5pt,belowskip=5pt,aboveskip=5pt]
plt.figure(num=None, figsize=(8, 4), dpi=600, facecolor='w', edgecolor='k')

plt.plot( x_values, [a-b for a,b in zip(y_values_old,y_values_true)], '-', color='#0000FF7F' )
plt.plot( x_values, [a-b for a,b in zip(y_values_new,y_values_true)], '-', color='#FF00007F' )

plt.legend( ["Numerical Recipes","Ours"] )
plt.xlabel("X")
plt.ylabel("Y")
plt.show()
\end{lstlisting}

\begin{figure*}[tbh]\begin{center}\adjustimage{max size={0.9\linewidth}{0.9\paperheight}}{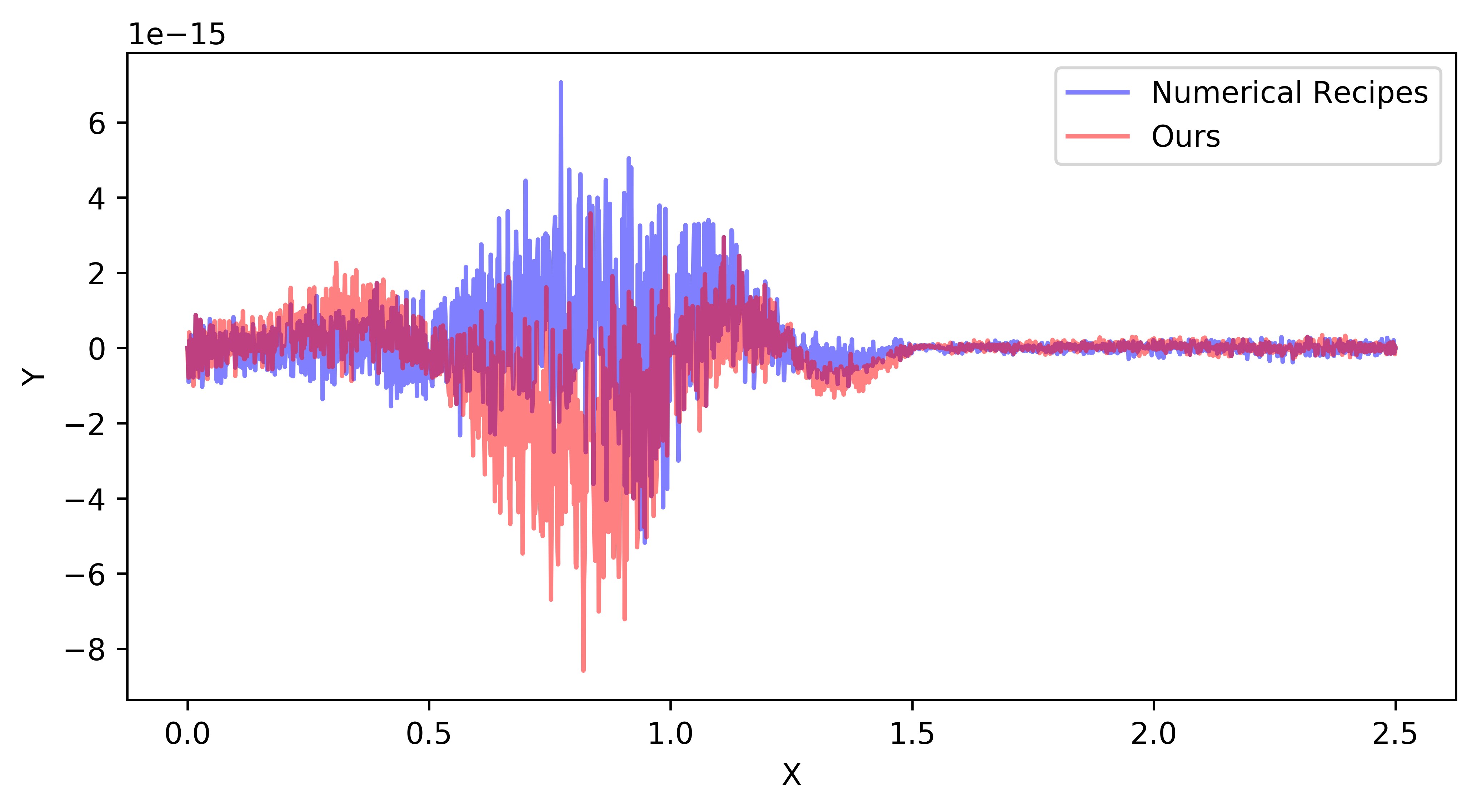}\end{center}\caption{Comparing the divergence between the routines and the ``true'' value.
Here we subtract the ``true'' value from the double-precision one.}\label{fig:diverge}\end{figure*}

\begin{lstlisting}[language=Python,numbers=none,xleftmargin=20pt,xrightmargin=5pt,belowskip=5pt,aboveskip=5pt]
table = pd.DataFrame( {'MSE, NR' :[mp.nstr(np.mean( [(a-b)**2 for a,b in zip(y_values_old,y_values_true)] ))], \
                       'MSE, our':[mp.nstr(np.mean( [(a-b)**2 for a,b in zip(y_values_new,y_values_true)] ))] })

display( {"text/latex":table.to_latex(index=False), "text/html":table.to_html(index=False)}, raw=True )
\end{lstlisting}

\begin{table}[H]
\caption{The mean square error of each set of routines from the ``true'' value.}\label{tbl:mse_full}
\centering
\begin{adjustbox}{max width=\textwidth}\rowcolors{2}{gray!20}{white}
\begin{tabular}{ll}
\toprule
     MSE, NR &    MSE, our \\
\midrule
 9.03518e-31 &  1.5724e-30 \\
\bottomrule
\end{tabular}

\end{adjustbox}
\end{table}

Overall, both Figure \ref{fig:diverge} and Table \ref{tbl:mse_full} show
that our code has less precision than the code of \emph{Numerical
Recipes}. The interactive Jupyter notebook agrees with this, though its
Figure \ref{fig:diverge} looks quite different.

\begin{lstlisting}[language=Python,numbers=none,xleftmargin=20pt,xrightmargin=5pt,belowskip=5pt,aboveskip=5pt]
nb_setup.images_hconcat(["fig_diverge_interactive.png"])
\end{lstlisting}

    \begin{figure*}[tbh]\begin{center}\adjustimage{max size={0.9\linewidth}{0.9\paperheight}}{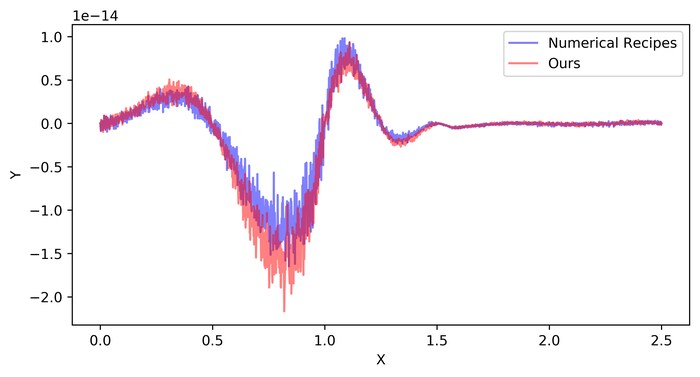}\end{center}\caption{Figure \ref{fig:diverge}, as seen in the Jupyter notebook used to
construct this paper.}\label{fig:diverge_int}\end{figure*}

It should be stressed that this imprecision is very small, on the order
of the precision expected for IEEE binary64 floating point numbers. This
level of imprecision can usually be safely ignored. It is also notable
that the mean square error has dropped as more points have been
introduced. This is further evidence neither algorithm has a systematic
bias from the true value.

\hypertarget{benchmarks}{%
\section{Benchmarks}\label{benchmarks}}

\texttt{nanoBench} benchmarks a short fragment of programming code by
repeatedly executing it multiple times in batches, while monitoring a
number of internal CPU counters.\cite{Abel2019nanoBenchAL} The default
measure used to summarize all batches is a trimmed mean that excludes
the top and bottom 20\% of samples, though users can select a median and
minimum instead. Both the trimmed mean and median assume execution times
are described by a Gaussian distribution. In theory, this cannot be the
case, as the Gaussian distribution has range \([-\infty,\infty]\) and
thus allows for negative times.

A more realistic model states that execution can be delayed via rare
events which occur with a fixed probability on any given clock tick.
This implies execution should follow a Poisson distribution, rather than
a Gaussian. For short code fragments where delays are unlikely this
Poisson distribution could resemble an Exponential distribution, while
for very long code fragments the Poisson may resemble a Gaussian
distribution more. The Poisson distribution can take a wide variety of
shapes, and no one metric of central tendency can hope to represent all
of them. This would be be less of a concern if \texttt{nanoBench}
displayed all three of the trimmed mean, median, and minimum, however
users can only select one of the three. Multiple measures require
multiple executions, or manual calculation based on the raw times
displayed in ``verbose'' mode.

\texttt{nanoBench} also lacks a good measure of variance, which can
provide vital information. Even \(O(1)\) code can have variable
execution time when given certain inputs. For instance, the authors have
observed one branchless series of floating-point operations that took
four times longer to execute when it was fed not-a-number or infinite
values. We have also seen that a code fragment which incorporated
division had more variance in execution times than one which did not. It
is entirely plausible for one code fragment to be 25\% faster on average
than a second that accomplishes the same task, yet the execution times
of the former exhibit so much variance that it only has a 51\% chance of
finishing before the latter. Knowledge of execution variance is
especially critical in real-time execution where code must finish before
a specific deadline.

Finally, \texttt{nanoBench} only tests a fixed number of repeats.
Testing a code fragment with \(O(n)\) behaviour requires manually
running it multiple times for various \(n\). This makes it difficult to
check that an algorithm does indeed have \(O(n)\) behaviour. Caches in
particular are notorious for exhibiting non-linear behaviour that could
bias performance measurements. By accident or design, a researcher could
select values of \(n\) which avoid non-linear cache behaviour for their
algorithm while triggering it in others' algorithms, distorting their
performance metrics.

We do not single out \texttt{nanoBench} because it is unusually flawed.
On the contrary, it has a better methodology than nearly all
benchmarking programs we have encountered. It shares a critical flaw of
nearly all benchmark methodologies within Computer Science: none of them
present enough information to assess their own fitness. This is
especially problematic for algorithms with run-times in the nanosecond
range, where even subtle sources of noise have a strong influence.

The ideal benchmarking methodology would instead attempt to execute the
algorithm over as much of the parameter space as possible, using a
process that minimizes systematic bias. Rather than assume a model, it
would either infer or assess a model against the raw data.

\hypertarget{curve-interpolation}{%
\subsection{Curve Interpolation}\label{curve-interpolation}}

The curve interpolation routines, \texttt{splint\_one} and
\texttt{newint}, are typically used on lists of control points. The most
common use is interpolation at fixed, regular distances along the curve
for display. As both algorithms are \(O(1)\), the typical approach to
benchmarking would generate a list of control points of a fixed size and
evaluate both routines. The runtime per evaluation is gathered by
clocking the time taken to process the entire list, then dividing by the
length of that list.

There is a hidden variable, however, the length of that list. If it is
fairly short, it may fit entirely within the L1 cache. If it is large,
it may spill into the L2 or L3 caches, if present, or main memory. All
of those memory pools have different access times, which in the case of
a fast algorithm may be enough to modify its expected runtime. Thus we
should treat both algorithms as if they were \(O(n)\) and use the
resulting dataset to either verify \(O(1)\) behaviour despite the caches
or extract their effect on performance.

While Python is an excellent prototyping language, users of
\texttt{newint} are likely to translate it into C or C++. We have thus
written our benchmarking program in C++.\cite{fast_cubic_source} For the
Intel i7-7700k, \texttt{gcc} 9.2.1 was used to compile the program. The
optimization flags \texttt{-O3\ -march=native} were used to maximize
performance, as it consistently generated faster code than \texttt{-O2}.

Each algorithm also comes in at least two variants. The original
\texttt{splint\_one} relied on division by a constant, which as
mentioned earlier was not converted to a multiplication by the compiler.
Thus we tested both a \texttt{splint\_one\_\_div}, which retained the
original division, as well as a \texttt{splint\_one\_\_mul} which
converted the division to multiplication by a constant. As for our
algorithms, \texttt{newint\_\_orig} corresponds exactly to
\texttt{newint}. We also tested a \texttt{newint\_\_noinv} variant that
eliminated the \texttt{inv\_ba} variable in favour of division by
\texttt{ba}. Finally, the \texttt{volatile} trick was tested via a
\texttt{newint\_\_vol} variant.

For all variations of \texttt{splint\_one} and \texttt{newint}, we
applied the following methodology.

\begin{enumerate}
\def\labelenumi{\arabic{enumi}.}
\tightlist
\item
  We executed all algorithms on curves with everywhere between four and
  1,048,576 control points.
\item
  The number of control points to use for any given run was contained in
  a list, which was generated by interpolation of \(0 \leq r \leq 1\)
  according to \((r(\sqrt{1,048,576}-\sqrt{4}) + \sqrt{4})^2\).
\item
  That list was generated by linearly interpolating \(r\) at a fixed
  number of points with constant spacing, totaling 524,288 points in
  most cases. Duplicates were not removed.
\item
  This list was shuffled into a random order.
\item
  Entries were drawn from that list to generate random control points of
  the form \(y_i = \epsilon\) and \(x_i = x_{i-1} + \epsilon\), where
  \(\epsilon \leftarrow U(0,1)\). If \(\epsilon = 0\), however, then
  \(x_i = x_{i-1} + 0.0001\). This ensures \(x_i > x_{i-1}\).
\item
  Second-order derivatives were generated for the control points using
  \texttt{new\_second\_derivative}.
\item
  A second list was generated with each draw, consisting of the order to
  execute the variants. This list was shuffled into a random order as
  well.
\item
  For each algorithm, a \texttt{for} loop generated uniformly-spaced
  points between \(x_1\) and \(x_n\) and interpolated the curve, taking
  advantage of the strictly increasing knots to simplify bracketing. One
  loop will not execute any of the algorithms but still calculate
  uniformly-spaced points and proper bracketing.
\item
  Execution was timed via C++'s \texttt{high\_resolution\_clock}.
\item
  These benchmarks were executed in a single thread.
\end{enumerate}

IEEE binary64 floating-point numbers were used. Large values of \(n\)
cause catastrophic precision loss for binary32 floating-point numbers.

The given formula was used instead of linear interpolation to create a
bias towards shorter list lengths, where those aforementioned secondary
factors would inject the most noise. As an intended side-effect, our
benchmarking code sampled some curve counts multiple times. For
instance, curves with four control points could be tested 1,376 times.
This over-sampling helps establish the statistical distribution of the
noisest execution regime.

The list of control point counts was randomized to ensure the computer
was not allowed to ``warm up'' to any given choice of control points. As
an intended side effect, should another process become active on the
same core it will show up as random noise in the data, provided the task
is only active for a short amount of time. If it is active for a
substantial time, a ``shadow'' of the true execution will show up in the
data. To ensure no algorithm had an advantage via cache warming, the
second list was introduced to randomly choose which algorithm to execute
first. The loop that does not run any algorithms will help estimate the
overhead of scanning over the control points and generating
uniformly-spaced points.

\texttt{glibc} aliases \texttt{high\_resolution\_clock} to
\texttt{system\_clock}, which on Linux provides approximately nanosecond
precision. While that clock does change as the system time is updated,
such updates occur with negligible chance and would be drowned out by
the amount of data collected. The use of a single thread is to minimize
the effects of hyperthreading and other processes executing on the
computer. Unless stated otherwise, these tests were executed on an idle
Intel i7-7700K processor, and the four cores it possesses should
minimize the effects of another process becoming active on the same
core.

As the benchmark is a C++ program that executes for hours, only the
output will be included here. The code is available online, though
\emph{Numerical Recipes}' code has been stripped out to remain in
complaince with its license.\cite{fast_cubic_source}

\begin{lstlisting}[language=Python,numbers=none,xleftmargin=20pt,xrightmargin=5pt,belowskip=5pt,aboveskip=5pt]
interp_bench = pd.read_csv("data/interp_range.i7-7700k.tsv", sep="\t" )

temp = interp_bench.copy()
temp.columns = [c.replace('_','\\_') for c in interp_bench.columns]
temp.tail()
\end{lstlisting}

\begin{table}[H]
\caption{The last few entries in the benchmark results. See the text for a full
explaination.}\label{tbl:intran_tail}
\centering
\begin{adjustbox}{max width=\textwidth}\rowcolors{2}{gray!20}{white}
\begin{tabular}{lrrrrrrrr}
\toprule
{} &       n &     noop &  splint\_one\_\_div &  splint\_one\_\_mul &  newint\_\_orig &  newint\_\_noinv &  newint\_\_vol &   order \\
\midrule
524283 &  474729 &  2629686 &             2918642 &             2938789 &         2940122 &          2884746 &        2860716 &   80480 \\
524284 &  832960 &  4579386 &             5136475 &             5183252 &         5148748 &          4988803 &        5092363 &   71492 \\
524285 &  105076 &   557215 &              630008 &              636847 &          632320 &           608602 &         619475 &    9003 \\
524286 &   30750 &   174446 &              185269 &              186036 &          185787 &           178701 &         181247 &  115221 \\
524287 &  687429 &  3840107 &             4227429 &             4305190 &         4267566 &          4138725 &        4158264 &  132329 \\
\bottomrule
\end{tabular}

\end{adjustbox}
\end{table}

All timing entries are in nanoseconds. \texttt{n} is the number of
control points in the curve, and \texttt{order} encodes the order of
execution of each algorithm. For the latter, the least three significant
bits specify the first algorithm run, the next three the second, and so
on. \texttt{noop} is the loop through the control points that calls none
of the above algorithms.

We can begin the benchmark analysis by graphing the timings for one of
the algorithms. One concern is whether or not the order of execution
matters.

\begin{lstlisting}[language=Python,numbers=none,xleftmargin=20pt,xrightmargin=5pt,belowskip=5pt,aboveskip=5pt]
# free up some memory
del temp
\end{lstlisting}

\begin{lstlisting}[language=Python,numbers=none,xleftmargin=20pt,xrightmargin=5pt,belowskip=5pt,aboveskip=5pt]
# the full dataset is too much clutter, instead pick a random subset
subset = interp_bench.sample( 64*1024 )
\end{lstlisting}

\begin{lstlisting}[language=Python,numbers=none,xleftmargin=20pt,xrightmargin=5pt,belowskip=5pt,aboveskip=5pt]
colors = ['#377eb8','#4daf4a','#984ea3','#e41a1c','#ff7f00']
plt.figure(num=None, figsize=(8, 4), dpi=600, facecolor='w', edgecolor='k')

for r in range(0,5):
    mask = (subset['order'] & (0x7 << 3*r)) == (3 << 3*r)
    plt.scatter( subset['n'][mask], (subset['newint__orig']/subset['n'])[mask], \
            marker='.', s=.03, color=colors[r], label="newint (original), round {}".format(r+1) )

plt.xlabel( "n" )
plt.ylabel( "nanoseconds / n" )

legend = plt.legend( loc='best' )
for i in range(0,5):
    legend.legendHandles[i]._sizes = [100]

plt.ylim( [5.8,6.6] )
plt.show()
\end{lstlisting}

\begin{figure*}[tbh]\begin{center}\adjustimage{max size={0.9\linewidth}{0.9\paperheight}}{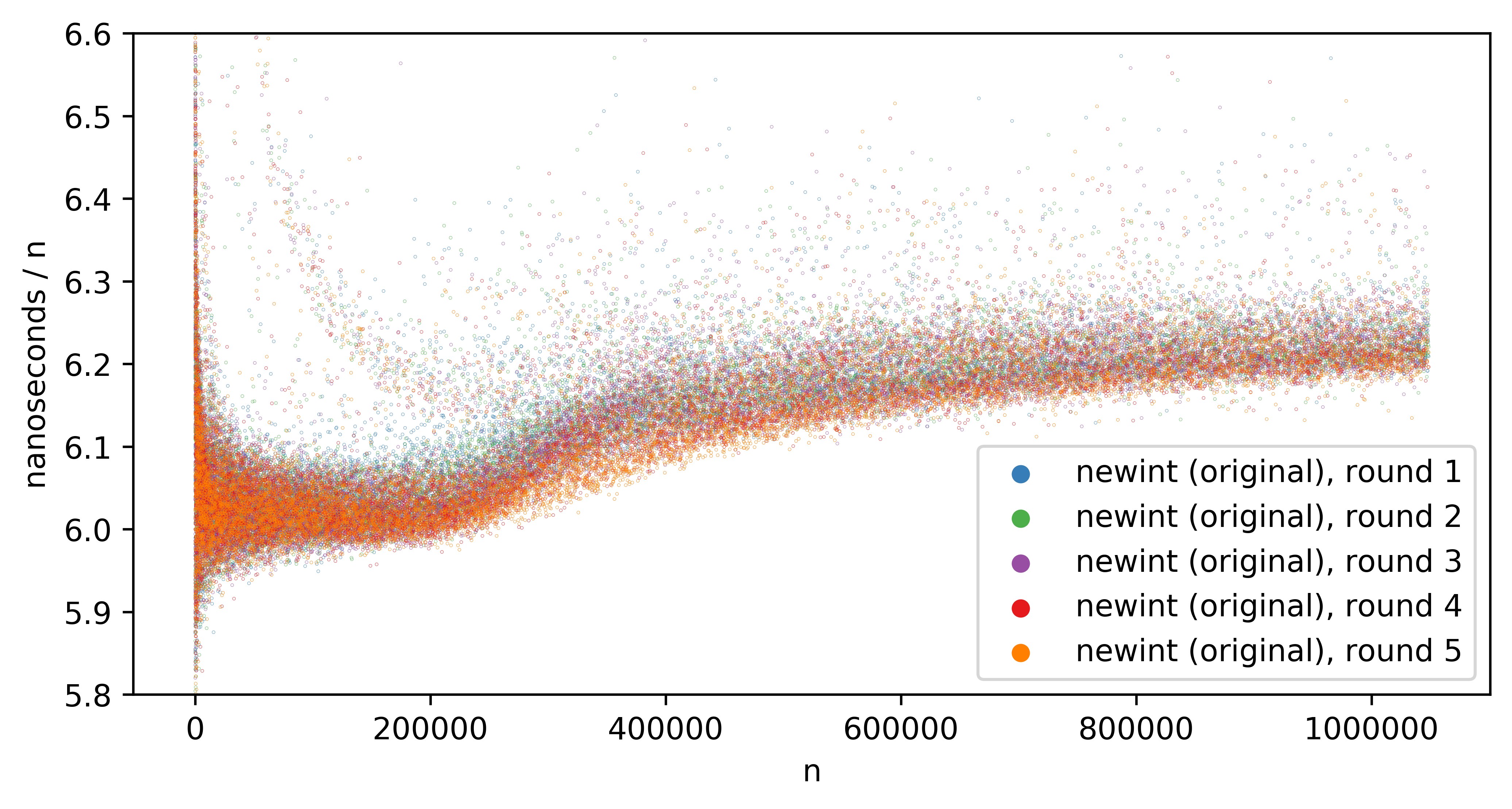}\end{center}\caption{Execution times per list length for our original interpolation
implementation.}\label{fig:newint_orig}\end{figure*}

Figure \ref{fig:newint_orig} shows the execution times of
\texttt{newint\_\_orig}. In general it shows no sign of achieving better
times if its execution comes after other algorithms. One notable
exception is the \(200,000 < n < 600,000\) range, where being executed
last provides a small but noticeable advantage. The dataset also shows a
distinctive ``goose neck'' starting at approximately \(n = 160,000\). A
possible explanation for both phenomenon relies on the L3 cache, which
for the Intel i7-7700k is 2MiB per core but shared across the four
cores. As there are three lists needed for the control points, and each
IEEE binary64 point within each list occupies eight bytes of memory,
2MiB can completely hold \(n \leq 87,381\) control points. Performance
begins to degrade when \(n\) is half the size of L3 and levels off when
\(n\) is twice the size of L3. The order-dependency between
\(200,000 < n < 600,000\) can be explained by the cache's pre-fetch
logic ``warming up'' to the data access patterns of the interpolation
task, and thus giving a minor performance boost to algorithms executed
later. Intelligent pre-fetching cannot help when the control point list
is entirely contained within L3, nor when the list is significantly
larger, so neither case shows order dependence.

Similar effects might be visible as the algorithm shifts from being
dominated by L1 fetches to L2 fetches, and from L2 to L3 fetch
dominance. The latter transition should begin at
\(5,461 \leq n \leq 10,922\), however there is a significant amount of
noise that could mask it and the former transition period between
\(682 \leq n \leq 1,365\) is even more obscured. The change in
memory-access latency may also be trivial next to the algorithm's
execution time, limiting the impact any performance increase could have.
Out-of-order execution could even eliminate the performance penalty, as
the processor could shuffle opcodes based on whether or not their memory
fetches have completed.

A secondary, slower performance profile is visible, following a Patero
distribution but merging with the primary execution profile at
approximately \(n = 300,000\). A third is faintly visible, with a
similar shape but slightly faster execution times. All execution orders
seem equally likely to appear, so the most likely explanation for these
profiles is contention with another process during execution.

The noise profile appears to follow a Poisson distribution. It is
dependent on \(n\) and scales with it. There is an increase in noise for
\(n < 50,000\) that is inversely proportional to \(n\), which implies a
second noise profile that does not scale with \(n\).

The ``goose neck'' can be explained by a Weibull distribution. This is
commonly found when describing a process with a failure rate \(\lambda\)
that changes according to a power of the current time, \(k\). Rather
than describing one process, though, consider a very large number of
them simultaneously. The odds of these processes failing is no longer
described by a probability density function, but instead the cumulative
distribution function of the Weibull distribution,

\begin{equation}
\text{CDF}_\text{Weibull}( x | k,\lambda ) = 1 - e^{ -\left( \frac{k}{\lambda}\right)^k }, \label{eqn:weibull_cdf}
\end{equation}

as the finite execution time limits how much of the PDF's domain could
be reached by any given process. If the Weibull distribution provides a
probabilistic model of cache misses within L3, this implies that the
replacement policy involves random replacement in some fashion. This is
in line with other research on Intel's cache
policies.\cite{abel_reverse_2014}

\begin{lstlisting}[language=Python,numbers=none,xleftmargin=20pt,xrightmargin=5pt,belowskip=5pt,aboveskip=5pt]
plt.figure(num=None, figsize=(8, 4), dpi=600, facecolor='w', edgecolor='k')

for r in range(0,5):
    mask = (subset['order'] & (0x7 << 3*r)) == (3 << 3*r)
    plt.scatter( subset['n'][mask], (subset['noop']/subset['n'])[mask], \
            marker='.', s=.03, color=colors[r], label="loop without any algorithm, round {}".format(r+1) )

plt.xlabel( "n" )
plt.ylabel( "nanoseconds / n" )

legend = plt.legend( loc='best' )
for i in range(0,5):
    legend.legendHandles[i]._sizes = [100]

plt.ylim( [4.8,6.6] )
plt.show()
\end{lstlisting}

\begin{figure*}[tbh]\begin{center}\adjustimage{max size={0.9\linewidth}{0.9\paperheight}}{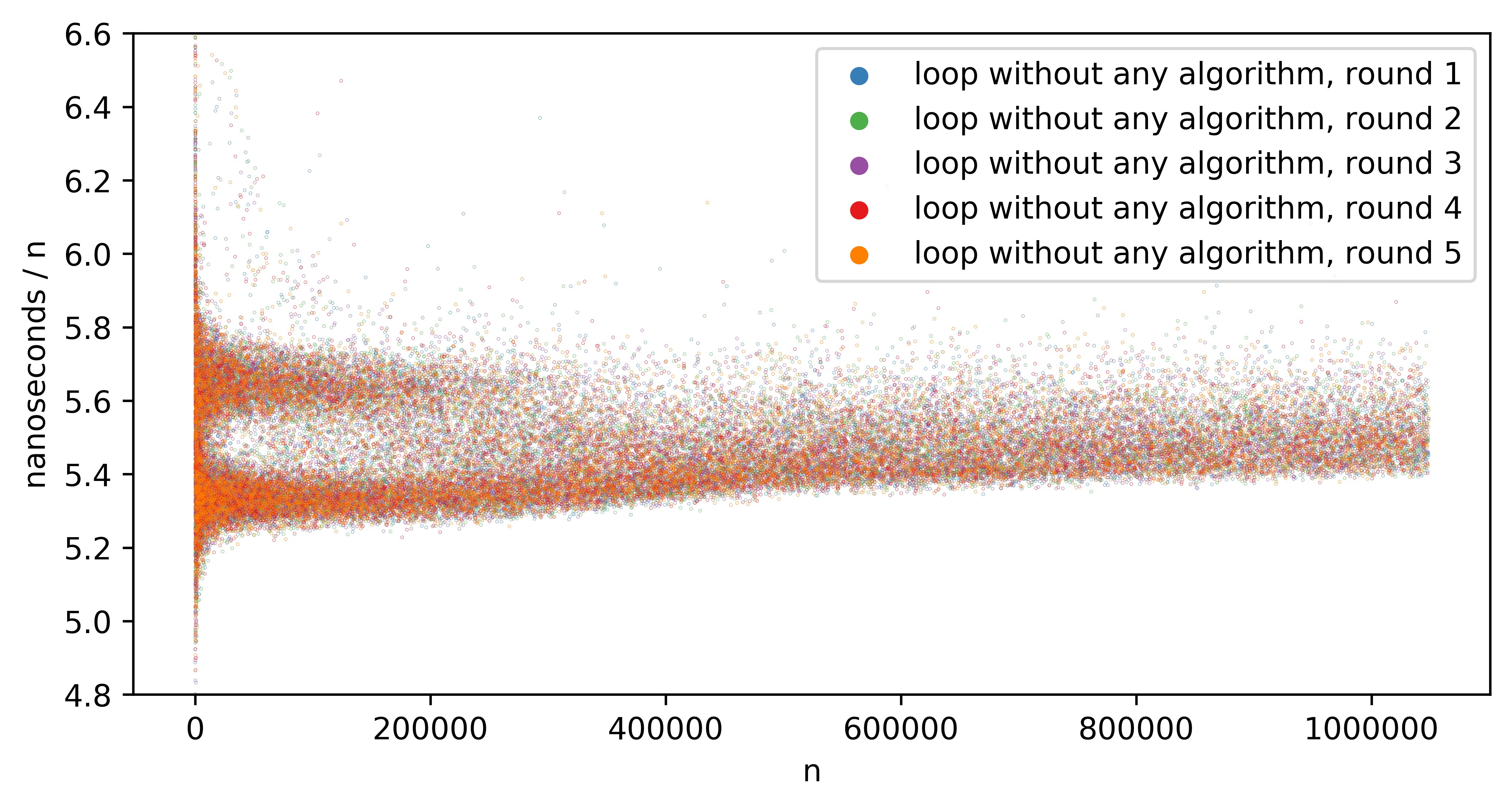}\end{center}\caption{Execution times per list length for `noop'.}\label{fig:interp_noop}\end{figure*}

Figure \ref{fig:interp_noop} shows the execution times when the
benchmarked code is run but no interpolation algorithm is executed. Two
execution paths are evident, with the slower of the two fading out
roughly when \(n\) is large enough to become main-memory dependent. No
order dependence is evident, surprisingly. Perhaps algorithm execution
is interfering with the cache policy logic.

The \texttt{noop} numbers make it clear that out-of-order execution
plays a major role in algorithm performance, at least on the i7-7700k.
If we estimate the mode of \texttt{noop}'s fastest execution path to be
5.3 ns/point, and the mode of \texttt{newint\_orig}'s slowest execution
point to be 6.2 ns/point, that leaves about one nanosecond per point to
execute the algorithm. Either the i7-7700k is executing two dozen
machine operations in about five clock cycles, or it is reordering
instructions so it can continue calculation while waiting for
memory-dependent operations to finish. It is also likely that register
renaming is allowing it to calculate multiple interpolants
simultaneously, further increasing performance.

\begin{lstlisting}[language=Python,numbers=none,xleftmargin=20pt,xrightmargin=5pt,belowskip=5pt,aboveskip=5pt]
# the algorithms tested, in order of their ID
algorithms = [
    'splint_one__div',
    'splint_one__mul',
    'newint__orig',
    'newint__noinv',
    'newint__vol'
            ]
\end{lstlisting}

\begin{lstlisting}[language=Python,numbers=none,xleftmargin=20pt,xrightmargin=5pt,belowskip=5pt,aboveskip=5pt]
plt.figure(num=None, figsize=(8, 4), dpi=600, facecolor='w', edgecolor='k')

for i,alg in enumerate(algorithms):

    # only include the middle execution
    mask = (subset['order'] & (0x7 << 3*2)) == ((i+1) << 3*2)
    plt.scatter( subset['n'][mask], (subset[alg]/subset['n'])[mask], alpha=0.3,
            marker='.', s=.1, color=colors[i], label="{}, round 3".format(alg) )

plt.ylabel( "nanoseconds / n" )

legend = plt.legend( loc='best' )
for i in range(0,5):
    legend.legendHandles[i]._sizes = [100]

plt.xlabel( "n" )
plt.ylabel( "nanoseconds / n" )

plt.ylim( [5.6,6.6] )

plt.show()
\end{lstlisting}

\begin{figure*}[tbh]\begin{center}\adjustimage{max size={0.9\linewidth}{0.9\paperheight}}{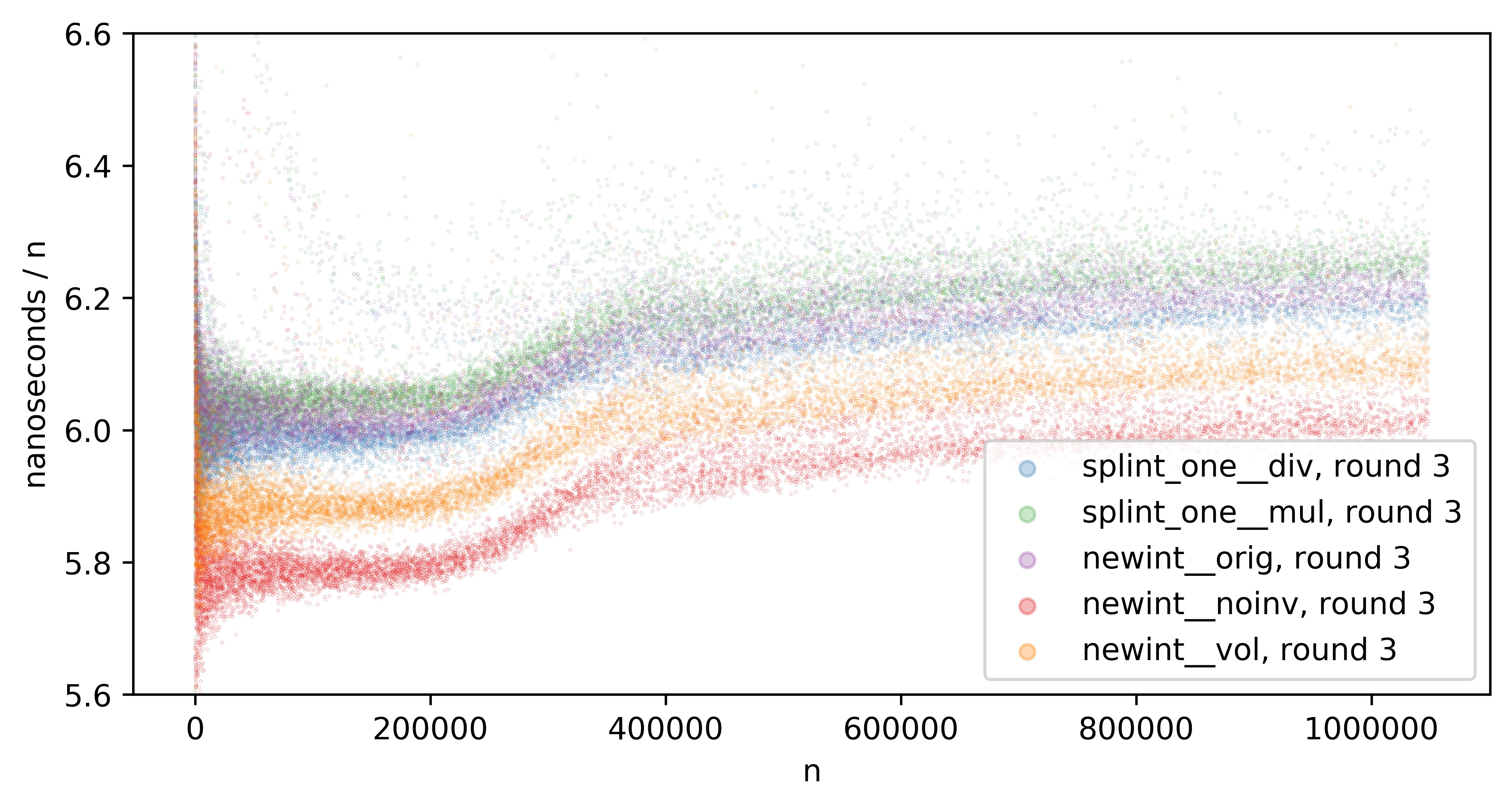}\end{center}\caption{Normalized execution times for all five algorithms.}\label{fig:exec_times_all}\end{figure*}

Figure \ref{fig:exec_times_all} shows the execution times of all
algorithms. All of them exhibit the same pattern of behaviour as
\texttt{newint\_\_orig}. This strongly implies the performance of all of
them can be captured by one model.

\hypertarget{execution-model}{%
\subsection{Execution Model}\label{execution-model}}

For this model, we can merge the two noise processes into one
distribution. The first key insight is that we do not need to rely on
the Poisson distribution.

\begin{align}
\text{Pois}( x | \lambda ) &= \frac{1}{x!} \lambda^x e^{-\lambda}, x \in \mathbb{Z} \label{eqn:poisson} \\
\Gamma( x | \kappa, \theta ) &= \frac{\theta^{-\kappa}}{\Gamma(\kappa)} x^{\kappa-1} e^{-\frac x \theta}, x \in \mathbb{R} \label{eqn:gamma}
\end{align}

Equation \ref{eqn:gamma} is the Gamma distribution, expressed in terms
of a shape parameter \(\kappa\) and a scale parameter \(\theta\). This
is not to be confused with the single-parameter \(\Gamma(x)\), which is
the Gamma function applied to x. Note that
\(\text{Pois}(x|\lambda) = \Gamma(\lambda|x+1,1)\), thus the Gamma
distribution is an analytic continuity of Equation \ref{eqn:poisson},
which describes the Poisson distribution. By using the Gamma
distribution, we do not have to concern ourselves with the quantized
nature of the Poisson distribution. The scale parameter exists only as a
convenience, as

\begin{equation}
\Gamma( x | \kappa, \theta ) = \Gamma( \frac x \theta | \kappa, 1 )
\end{equation}

The second key insight is that both the mode and standard deviation of
any given Gamma distribution have a simple closed form, provided
\(\kappa \geq 1\).

\begin{lstlisting}[language=Python,numbers=none,xleftmargin=20pt,xrightmargin=5pt,belowskip=5pt,aboveskip=5pt]
mode, sigma, kappa, theta = sp.symbols('Mo \\sigma \\kappa \\theta', real=True, positive=True)

first  = sp.Eq( mode, (kappa - 1)*theta )
peq( first )
\end{lstlisting}

\begin{equation}\label{eqn:gamma_mode}
Mo = \theta \left(\kappa - 1\right)
\end{equation}

\begin{lstlisting}[language=Python,numbers=none,xleftmargin=20pt,xrightmargin=5pt,belowskip=5pt,aboveskip=5pt]
second = sp.Eq( sigma, theta*sp.sqrt(kappa) )
peq( second )
\end{lstlisting}

\begin{equation}\label{eqn:gamma_sigma}
\sigma = \sqrt{\kappa} \theta
\end{equation}

We can rearrange Equations \ref{eqn:gamma_mode} and
\ref{eqn:gamma_sigma} to reparameterize the Gamma distribution, so it
instead takes a mode and standard deviation as input.

\begin{lstlisting}[language=Python,numbers=none,xleftmargin=20pt,xrightmargin=5pt,belowskip=5pt,aboveskip=5pt]
# solve the first equation for kappa
temp = sp.solve( first, kappa )[0]

# substitute this into the second equation
temp = second.subs( kappa, temp )

# solve for theta
third = sp.Eq( theta, sp.solve( temp, theta )[0] )

peq( third )
\end{lstlisting}

\begin{equation}\label{eqn:gamma_theta}
\theta = - \frac{Mo}{2} + \frac{\sqrt{Mo^{2} + 4 \sigma^{2}}}{2}
\end{equation}

\begin{lstlisting}[language=Python,numbers=none,xleftmargin=20pt,xrightmargin=5pt,belowskip=5pt,aboveskip=5pt]
# substitute back into the first equation
temp = first.subs( theta, third.rhs )

# solve for kappa
fourth = sp.Eq( kappa, sp.solve( temp, kappa )[0] )

peq( fourth )
\end{lstlisting}

\begin{equation}\label{eqn:gamma_kappa}
\kappa = - \frac{Mo + \sqrt{Mo^{2} + 4 \sigma^{2}}}{Mo - \sqrt{Mo^{2} + 4 \sigma^{2}}}
\end{equation}

Equations \ref{eqn:gamma_kappa} and \ref{eqn:gamma_theta} allow us to
combine both noise distributions into one. We hold the mode to be
constant, and vary the standard deviation according to

\begin{equation}
\sigma = n \cdot \sigma_{\infty} + \sigma_0 \label{eqn:model_noise_stdev}
\end{equation}

That mode constitutes the offset above a baseline execution time, the
fastest the algorithm could be executed on the given processor in
theory. As Figure \ref{fig:newint_orig} demonstrates, there are two
primary execution profiles, one where access to the L3 cache dominates
performance, and one where access to main memory dominates. These both
are \(O(n)\), which means both can be represented as a linear relation
on \(n\). After some transition point \(t\), the execution time
transitions between the L3-dominated profile to the main-memory
dominated profile according to the culmulative Weibull distribution. We
can express this baseline as

\begin{align}
\text{baseline} &= w(n \cdot m_{L3} + b_{L3}) + (1-w)(n \cdot m_{MM}), \label{eqn:model_base} \\
w &= \begin{cases}
e^{ -\left( (n-t)s \right)^p },&~ n > t \\
1,&~ \text{otherwise.}
\end{cases}
\end{align}

where \(m_{L3}\) and \(m_{MM}\) are the slope of the L3-dependent and
main-memory-dependent execution profiles, plus \(s\) and \(p\) are
scaling and power factors for the Weibull distribution. The intercept
for the main-memory-dependent profile was dropped as the execution times
are sufficiently large that it would have negligible influence.

Those large execution times pose a pratical problem, as floating-point
numbers only offer a finite precision. To remove that problem, we divide
both the exectution times and Equations \ref{eqn:model_noise_stdev} and
\ref{eqn:model_base} by \(n\). This also makes the model easier to fit.

\begin{align}
p( y | n, m_{L3}, b_{L3}, m_{MM}, t, s, p, \textit{Mo}, \sigma_0, \sigma_\infty ) &= \Gamma( \frac y n - \text{baseline} | \textit{Mo}, \sigma ) \\
\sigma &= \sigma_{\infty} + \frac{\sigma_0}{n} \\
\text{baseline} &= w(m_{L3} + \frac{b_{L3}}{n}) + (1-w)m_{MM} \\
w &= \begin{cases}
e^{ -\left( (n-t)s \right)^p },&~ n > t \\
1,&~ \text{otherwise.}
\end{cases}
\end{align}

We used \texttt{emcee} to fit this model to the
data.\cite{foreman-mackey_emcee:_2013} \texttt{PyMC3} would have been a
more natural choice,\cite{Salvatier2016} but it ran into difficulties
generating a derivative. While the former is likely slower to converge
on the posterior than the latter, it relies on pure Python functions
which are easier to include in this paper.

As a full round of MCMC can take hours, we will not include the
programming code here. Those wishing to check our work are encouraged to
view the source code repository, where the notebooks used to test
various models are stored alongside the
datasets.\cite{fast_cubic_source} These also include model checks, to
ensure a goodness of fit. For now, we will simply load an archived copy
of the resulting posteriors and chart the fitted values.

\begin{lstlisting}[language=Python,numbers=none,xleftmargin=20pt,xrightmargin=5pt,belowskip=5pt,aboveskip=5pt]
posteriors = list()
for alg in algorithms:
    posteriors.append( np.loadtxt("posterior.{}.interp_range.i7-7700k.variable_gamma.tsv".format(alg),
                              delimiter='\t') )
    np.random.shuffle( posteriors[-1] )   # shuffle to destroy any order

print( "There are {} samples within each posterior.".format( len(posteriors[0]) ) )
\end{lstlisting}

\begin{lstlisting}[language={},postbreak={},numbers=none,xrightmargin=7pt,belowskip=5pt,aboveskip=5pt,breakindent=0pt]
There are 4096 samples within each posterior.

\end{lstlisting}

\begin{lstlisting}[language=Python,numbers=none,xleftmargin=20pt,xrightmargin=5pt,belowskip=5pt,aboveskip=5pt]
table = pd.DataFrame( {'variable':['$m_{L3}$', '$m_{MM}$', '$b_{L3}$',
                                   '$t$', '$s$', '$p$',
                                   '$\textit{Mo}$', '$\sigma_0$', '$\sigma_\infty$']} )

for i,alg in enumerate(algorithms):

    # https://emcee.readthedocs.io/en/v2.2.1/user/line/#results
    table[alg.replace("_","\\_")] = list(map( lambda v: \
                    "${{ {:.2e}^{{+ {:.2e}}}_{{- {:.2e}}} }}$".format(v[1], v[2]-v[1], v[1]-v[0]),
                    zip(*np.percentile(posteriors[i], [16, 50, 84], axis=0)) ))

display( {"text/latex":table.to_latex(index=False), "text/html":table.to_html(index=False)}, raw=True )
\end{lstlisting}

\begin{table}[H]
\caption{The posterior distribution for each variable of the model, including
16/84 credible intervals.}\label{tbl:vargamma_posteriors}
\centering
\begin{adjustbox}{max width=\textwidth}\rowcolors{2}{gray!20}{white}
\begin{tabular}{llllll}
\toprule
        variable &                        splint\_one\_\_div &                        splint\_one\_\_mul &                            newint\_\_orig &                           newint\_\_noinv &                             newint\_\_vol \\
\midrule
        $m_{L3}$ &  ${ 5.32e+00^{+ 4.37e-05}_{- 8.25e-05} }$ &  ${ 5.25e+00^{+ 7.72e-05}_{- 1.33e-04} }$ &  ${ 5.28e+00^{+ 9.03e-05}_{- 4.29e-03} }$ &  ${ 4.90e+00^{+ 1.03e-04}_{- 1.61e-04} }$ &  ${ 5.16e+00^{+ 5.96e-05}_{- 9.68e-05} }$ \\
        $m_{MM}$ &  ${ 5.53e+00^{+ 1.45e-03}_{- 1.70e-03} }$ &  ${ 5.46e+00^{+ 1.31e-03}_{- 1.54e-03} }$ &  ${ 5.49e+00^{+ 1.32e-01}_{- 2.44e-03} }$ &  ${ 5.14e+00^{+ 1.67e-03}_{- 1.23e-03} }$ &  ${ 5.38e+00^{+ 1.40e-03}_{- 1.43e-03} }$ \\
        $b_{L3}$ &  ${ 2.39e-03^{+ 3.82e-03}_{- 1.84e-03} }$ &  ${ 2.04e-03^{+ 4.55e-03}_{- 1.55e-03} }$ &  ${ 3.05e-03^{+ 6.59e-03}_{- 2.41e-03} }$ &  ${ 8.27e-03^{+ 1.53e-02}_{- 6.09e-03} }$ &  ${ 1.63e-03^{+ 2.81e-03}_{- 1.21e-03} }$ \\
             $t$ &  ${ 2.18e+05^{+ 2.23e+03}_{- 2.10e+03} }$ &  ${ 2.19e+05^{+ 1.35e+03}_{- 1.87e+03} }$ &  ${ 2.37e+05^{+ 1.23e+04}_{- 2.67e+03} }$ &  ${ 2.17e+05^{+ 2.49e+03}_{- 1.12e+03} }$ &  ${ 2.20e+05^{+ 1.91e+03}_{- 1.62e+03} }$ \\
             $s$ &  ${ 5.63e-06^{+ 9.24e-08}_{- 1.02e-07} }$ &  ${ 5.33e-06^{+ 8.02e-08}_{- 9.33e-08} }$ &  ${ 5.53e-06^{+ 1.53e-07}_{- 4.61e-06} }$ &  ${ 5.47e-06^{+ 8.53e-08}_{- 8.62e-08} }$ &  ${ 5.68e-06^{+ 8.47e-08}_{- 9.45e-08} }$ \\
             $p$ &  ${ 8.03e-01^{+ 2.09e-02}_{- 2.04e-02} }$ &  ${ 7.91e-01^{+ 1.98e-02}_{- 1.45e-02} }$ &  ${ 6.85e-01^{+ 2.50e-02}_{- 2.74e-01} }$ &  ${ 8.09e-01^{+ 1.46e-02}_{- 1.86e-02} }$ &  ${ 7.96e-01^{+ 1.84e-02}_{- 1.81e-02} }$ \\
   $\textit{Mo}$ &  ${ 6.67e-01^{+ 3.86e-04}_{- 3.98e-04} }$ &  ${ 8.03e-01^{+ 3.41e-04}_{- 3.20e-04} }$ &  ${ 7.53e-01^{+ 2.14e-03}_{- 4.03e-04} }$ &  ${ 8.93e-01^{+ 3.98e-04}_{- 4.07e-04} }$ &  ${ 7.29e-01^{+ 3.56e-04}_{- 3.80e-04} }$ \\
      $\sigma_0$ &  ${ 4.81e+02^{+ 7.26e+00}_{- 7.70e+00} }$ &  ${ 6.46e+02^{+ 1.10e+01}_{- 9.92e+00} }$ &  ${ 5.90e+02^{+ 1.07e+01}_{- 1.95e+01} }$ &  ${ 6.01e+02^{+ 1.02e+01}_{- 8.85e+00} }$ &  ${ 7.87e+02^{+ 1.10e+01}_{- 9.17e+00} }$ \\
 $\sigma_\infty$ &  ${ 7.95e-02^{+ 2.13e-04}_{- 2.61e-04} }$ &  ${ 7.59e-02^{+ 2.28e-04}_{- 2.32e-04} }$ &  ${ 7.68e-02^{+ 1.74e-03}_{- 2.79e-04} }$ &  ${ 7.42e-02^{+ 2.33e-04}_{- 2.21e-04} }$ &  ${ 7.23e-02^{+ 2.57e-04}_{- 2.14e-04} }$ \\
\bottomrule
\end{tabular}

\end{adjustbox}
\end{table}

One surprise is that the fitted Weibull power is below one for every
algorithm, leading to a sharp transition point instead of the smooth
``goose neck'' transition observed in the data. One possible explanation
is that the Weibull distribution assumes the failure rate changes
according to a constant power of the time. The failure rate may instead
change according to a variable power of the time. Another is that there
are multiple execution profiles with different parameters for each,
rather than just one, and the fitted power was the best compromise among
all of them. Whatever the explanation, the culmulative Weibull portion
of the model is the least certain portion of the model.

\begin{lstlisting}[language=Python,numbers=none,xleftmargin=20pt,xrightmargin=5pt,belowskip=5pt,aboveskip=5pt]
columns = ['n','count'] + [s.replace("_","\\_") for s in algorithms]
data = [list() for i in columns]

for n in [4, 8, 16, 32, 64, 128]:
    aggregated = interp_bench[ interp_bench['n'] == n ].drop(columns=['order','noop']).groupby(by='n'
            ).agg( [lambda x: np.percentile(x/n, 16),
                    lambda x: np.median(x/n),
                    lambda x: np.percentile(x/n, 84),
                    len] ).reset_index()

    data[0].append(n)
    data[1].append(aggregated[('newint__vol','len')].iloc[0])

    for i,alg in enumerate(algorithms):
        data[i+2].append( "${{ {:.2f}^{{+ {:.2f}}}_{{- {:.2f}}} }}$".format(
            aggregated[(alg,'<lambda_1>')].iloc[0],
            aggregated[(alg,'<lambda_2>')].iloc[0] - aggregated[(alg,'<lambda_1>')].iloc[0],
            aggregated[(alg,'<lambda_1>')].iloc[0] - aggregated[(alg,'<lambda_0>')].iloc[0] ) )

    del aggregated

table = pd.DataFrame( {columns[i]:data[i] for i,_ in enumerate(data)} )
display( {"text/latex":table.to_latex(index=False), "text/html":table.to_html(index=False)}, raw=True )
\end{lstlisting}

\begin{table}[H]
\caption{Calculated medians and 16/84 percentiles for select small values of n.}\label{tbl:small_n_fail}
\centering
\begin{adjustbox}{max width=\textwidth}\rowcolors{2}{gray!20}{white}
\begin{tabular}{rrlllll}
\toprule
   n &  count &             splint\_one\_\_div &             splint\_one\_\_mul &                 newint\_\_orig &                newint\_\_noinv &                   newint\_\_vol \\
\midrule
   4 &    122 &  ${ 13.00^{+ 2.32}_{- 1.50} }$ &  ${ 14.50^{+ 2.25}_{- 2.25} }$ &  ${ 12.50^{+ 6.81}_{- 2.50} }$ &  ${ 12.00^{+ 3.73}_{- 1.50} }$ &  ${ 13.50^{+ 13.25}_{- 3.00} }$ \\
   8 &     89 &   ${ 9.88^{+ 1.62}_{- 1.11} }$ &  ${ 10.50^{+ 1.85}_{- 1.49} }$ &   ${ 9.38^{+ 1.88}_{- 1.12} }$ &   ${ 9.88^{+ 2.73}_{- 1.50} }$ &    ${ 9.88^{+ 5.74}_{- 1.50} }$ \\
  16 &     63 &   ${ 8.19^{+ 1.11}_{- 1.06} }$ &   ${ 8.38^{+ 1.00}_{- 0.82} }$ &   ${ 8.06^{+ 1.20}_{- 1.44} }$ &   ${ 7.62^{+ 1.52}_{- 1.00} }$ &    ${ 7.88^{+ 2.67}_{- 0.82} }$ \\
  32 &     45 &   ${ 7.31^{+ 0.53}_{- 0.56} }$ &   ${ 7.44^{+ 1.29}_{- 0.84} }$ &   ${ 7.16^{+ 1.00}_{- 0.90} }$ &   ${ 7.03^{+ 0.99}_{- 0.91} }$ &    ${ 7.56^{+ 1.39}_{- 1.09} }$ \\
  64 &     31 &   ${ 6.84^{+ 0.53}_{- 0.36} }$ &   ${ 6.88^{+ 0.45}_{- 0.42} }$ &   ${ 6.80^{+ 0.82}_{- 0.36} }$ &   ${ 6.80^{+ 0.47}_{- 0.52} }$ &    ${ 6.80^{+ 0.83}_{- 0.54} }$ \\
 128 &     23 &   ${ 6.46^{+ 0.62}_{- 0.41} }$ &   ${ 6.47^{+ 0.62}_{- 0.33} }$ &   ${ 6.53^{+ 0.29}_{- 0.41} }$ &   ${ 6.29^{+ 0.28}_{- 0.55} }$ &    ${ 6.48^{+ 0.85}_{- 0.57} }$ \\
\bottomrule
\end{tabular}

\end{adjustbox}
\end{table}

Of all the variables, \(b_{L3}\) shows the most uncertainty within the
model and is on the order of a few picoseconds. Combined with a
consistently large \(\sigma_0\) that exhibits little uncertainty, this
would suggest \(b_{L3}\) contributes little to the overall fit. As Table
\ref{tbl:small_n_fail} demonstrates, however, there is an increase in
the median execution time per \(n\) as \(n\) approaches zero, which
argues an offset variable like \(b_{L3}\) should play a significant role
in any model. The output of this model for small \(n\) should not be
trusted.

\begin{lstlisting}[language=Python,numbers=none,xleftmargin=20pt,xrightmargin=5pt,belowskip=5pt,aboveskip=5pt]
import scipy.stats as sps

def gamma_param( mode, sigma ):
    """Convert the given mode and standard deviation into a Gamma shape and scale parameter."""

    root     = np.sqrt( mode*mode + 2*sigma*sigma )         # simplify calculation
    rootmode = root - mode

    theta = 0.5*rootmode
    kappa = (mode + root) / rootmode

    return (kappa, theta)

def model_bilinear( theta, n ):
    """Calcaluate the model's baseline."""

    m_l3,m_mm,b_l3, t_l3_mm,w_s,w_p, mode,sigma_0,sigma_infty = theta

    model = m_l3 + b_l3/n

    # early exit, if possible
    if ((type(n) is np.int64) or (type(n) is int)) and (n < t_l3_mm):
            return model

    # use clipping to ensure sane values
    frac = np.clip( np.nan_to_num(np.exp( -( (n - t_l3_mm)*w_s )**w_p ), nan=1.), 0., 1. )

    model = model*frac + m_mm*(1 - frac)
    return model

def model_noise_stdev( theta, n ):
    """Calculate the noise standard deviation, given n."""

    m_l3,m_mm,b_l3, t_l3_mm,w_s,w_p, mode,sigma_0,sigma_infty = theta

    return sigma_infty + sigma_0/n

def model_range_vargamma( theta, n, fraction=(2./3.) ):
    """Determine the credible interval for the given parameter set."""

    m_l3,m_mm,b_l3, t_l3_mm,w_s,w_p, mode,sigma_0,sigma_infty = theta

    baseline = model_bilinear(theta, n)

    model_stdev      = model_noise_stdev(theta, n)
    model_kappa, model_theta = gamma_param( mode, model_stdev )

    low, high = sps.gamma.interval( fraction, model_kappa, 0, model_theta )
    return (baseline + low, baseline + high)
\end{lstlisting}

\begin{lstlisting}[language=Python,numbers=none,xleftmargin=20pt,xrightmargin=5pt,belowskip=5pt,aboveskip=5pt]
plt.figure(num=None, figsize=(8, 4), dpi=600, facecolor='w', edgecolor='k')

n = np.linspace( np.min(subset['n']), np.max(subset['n']), 512 )

# background: 16/84 confidence interval
for theta in posteriors[2][0:50]:

        m_l3,m_mm,b_l3, t_l3_mm,w_s,w_p, mode,sigma_0,sigma_infty = theta

        lower, upper = model_range_vargamma( theta, n )

        plt.fill_between( n, lower, upper, facecolor=colors[2], alpha=0.002 )

# middle: the datapoints themselves
mask = (subset['order'] & (0x7 << 3*2)) == (3 << 3*2)
plt.scatter( subset['n'][mask], (subset['newint__orig']/subset['n'])[mask], alpha=1,
            marker='.', s=.03, color=colors[2], label="newint (original), round 3" )

# foreground: the adjusted mode
for theta in posteriors[2][0:100]:

        m_l3,m_mm,b_l3, t_l3_mm,w_s,w_p, mode,sigma_0,sigma_infty = theta

        adj_mode     = model_bilinear( theta, n ) + mode
        plt.plot( n, adj_mode, color=colors[2], alpha=0.02 )

plt.xlabel( "n" )
plt.ylabel( "nanoseconds / n" )

legend = plt.legend( loc='best' )
legend.legendHandles[0]._sizes = [100]

plt.ylim( [5.8,6.6] )

plt.show()
\end{lstlisting}

\begin{lstlisting}[language={},postbreak={},numbers=none,xrightmargin=7pt,belowskip=5pt,aboveskip=5pt,breakindent=0pt]
/home/hjhornbeck/.local/lib/python3.7/site-packages/ipykernel_launcher.py:26: RuntimeWarning: invalid value encountered in power

\end{lstlisting}

\begin{figure*}[tbh]\begin{center}\adjustimage{max size={0.9\linewidth}{0.9\paperheight}}{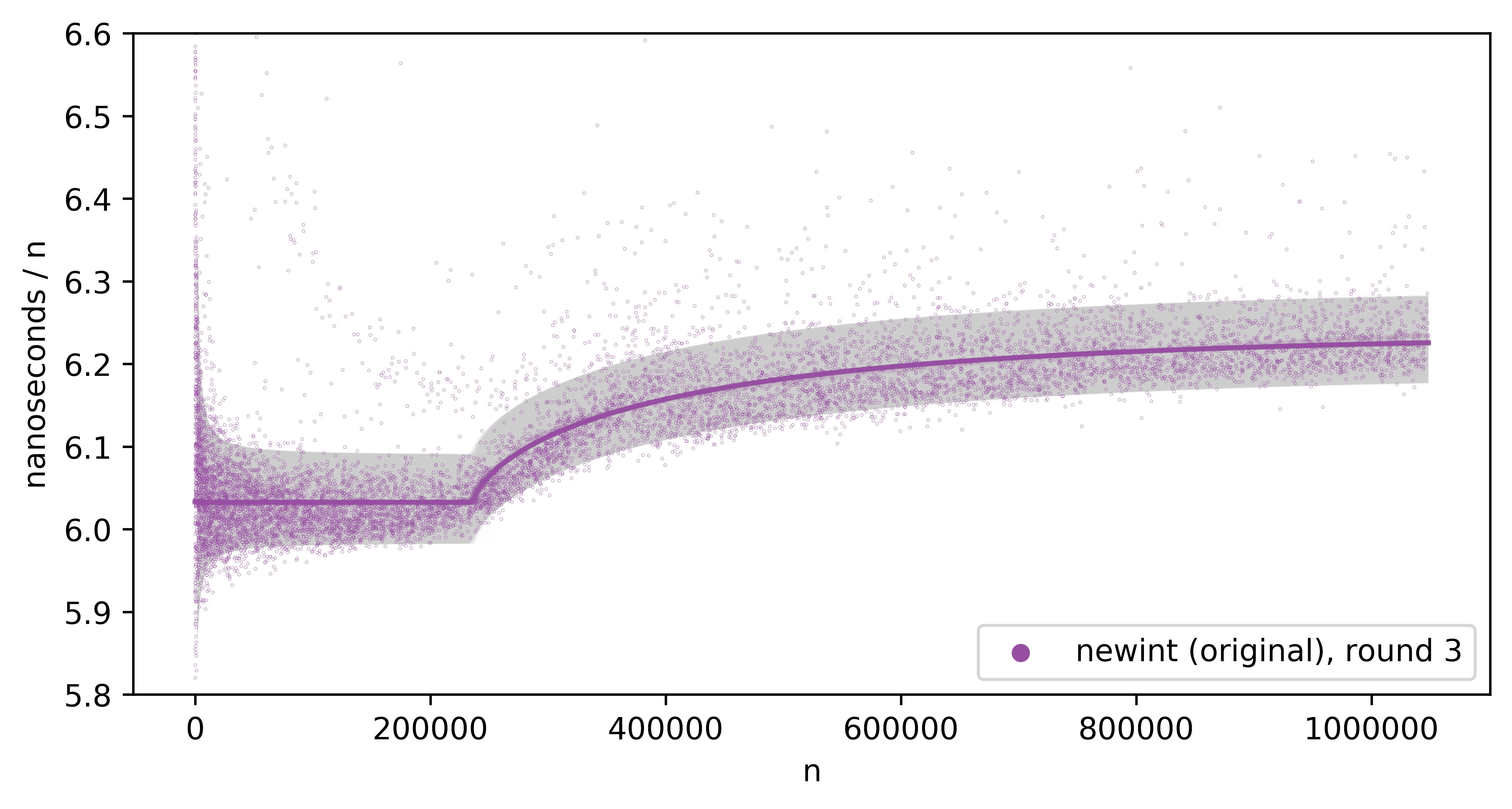}\end{center}\caption{Execution times per list length for `newint', original version, for when
it is run third among the algorithms. A sample of the model's posterior
is displayed for comparison. The solid line is the modal execution time,
while the 16/84 confidence interval is the filled region.}\label{fig:newint_model_comp}\end{figure*}

In comparison, Figure \ref{fig:newint_model_comp} suggests the model is
a good fit overall. We can resolve this contradiction by randomly
drawing sampling from the posterior, and evaluating the likelihood of a
subset of datapoints.

\begin{lstlisting}[language=Python,numbers=none,xleftmargin=20pt,xrightmargin=5pt,belowskip=5pt,aboveskip=5pt]
def lnprior_vargamma(theta):
    """Calculate the prior probability of the given parameters."""

    m_l3,m_mm,b_l3, t_l3_mm,w_s,w_p, mode,sigma_0,sigma_infty = theta

    if (m_l3 <= 0) or (m_mm <= 0):
        return -np.inf
    if m_l3 > m_mm:                               # L3 MUST be faster than main memory
        return -np.inf
    if b_l3 < 0:
        return -np.inf

    if (t_l3_mm < 32000) or (t_l3_mm > 350000):   # i7-7700k
        return -np.inf
    if (w_s > 1e-2) or (w_s <= 0):
        return -np.inf
    if (w_p <= 0) or (w_p > 1000):                # i7-7700k
        return -np.inf

    if (mode <= 0) or (sigma_0 <= 0) or (sigma_infty <= 0):
        return -np.inf
    if (sigma_infty > sigma_0):                        # we know the noise decreases
        return -np.inf

    return -1.5 * (np.log1p( m_l3*m_l3 ) + np.log1p( m_mm*m_mm ))

def lnlike_vargamma(theta, x, y):
    """Calculate the likelihood of the observed y, given the dependent x and a set of parameters."""

    m_l3,m_mm,b_l3, t_l3_mm,w_s,w_p, mode,sigma_0,sigma_infty = theta

    normed      = y / x
    model_floor = model_bilinear( theta, x )

    model_stdev = model_noise_stdev( theta, x )
    model_kappa, model_theta = gamma_param( mode, model_stdev )

    return np.sum( sps.gamma.logpdf( normed, model_kappa, model_floor, model_theta ) )
\end{lstlisting}

\begin{lstlisting}[language=Python,numbers=none,xleftmargin=20pt,xrightmargin=5pt,belowskip=5pt,aboveskip=5pt]
plt.figure(num=None, figsize=(8, 4), dpi=600, facecolor='w', edgecolor='k')

total_points     = 64*1024
total_posteriors = 128
points_per_posterior = total_points // total_posteriors

# only take the middle execution, as well as values of n below a threshold
mask = ((interp_bench['order'] & (0x7 << 3*2)) == (3 << 3*2)) & (interp_bench['n'] < 300000)

for theta in posteriors[2][0:total_posteriors]:
    subsubset = interp_bench[ mask ].sample( points_per_posterior )

    x_subset = subsubset['n']
    y_subset = subsubset['newint__orig']

    llike    = [lnlike_vargamma(theta, x_subset.iloc[i], y_subset.iloc[i]) for i in range(len(x_subset))]

    plt.scatter( x_subset, llike, alpha=0.3, marker='.', s=.1, color=colors[2] )

plt.xlabel( "n" )
plt.ylabel( "log(likelihood)" )

plt.ylim( [-5,2.2] )

plt.show()
\end{lstlisting}

\begin{figure*}[tbh]\begin{center}\adjustimage{max size={0.9\linewidth}{0.9\paperheight}}{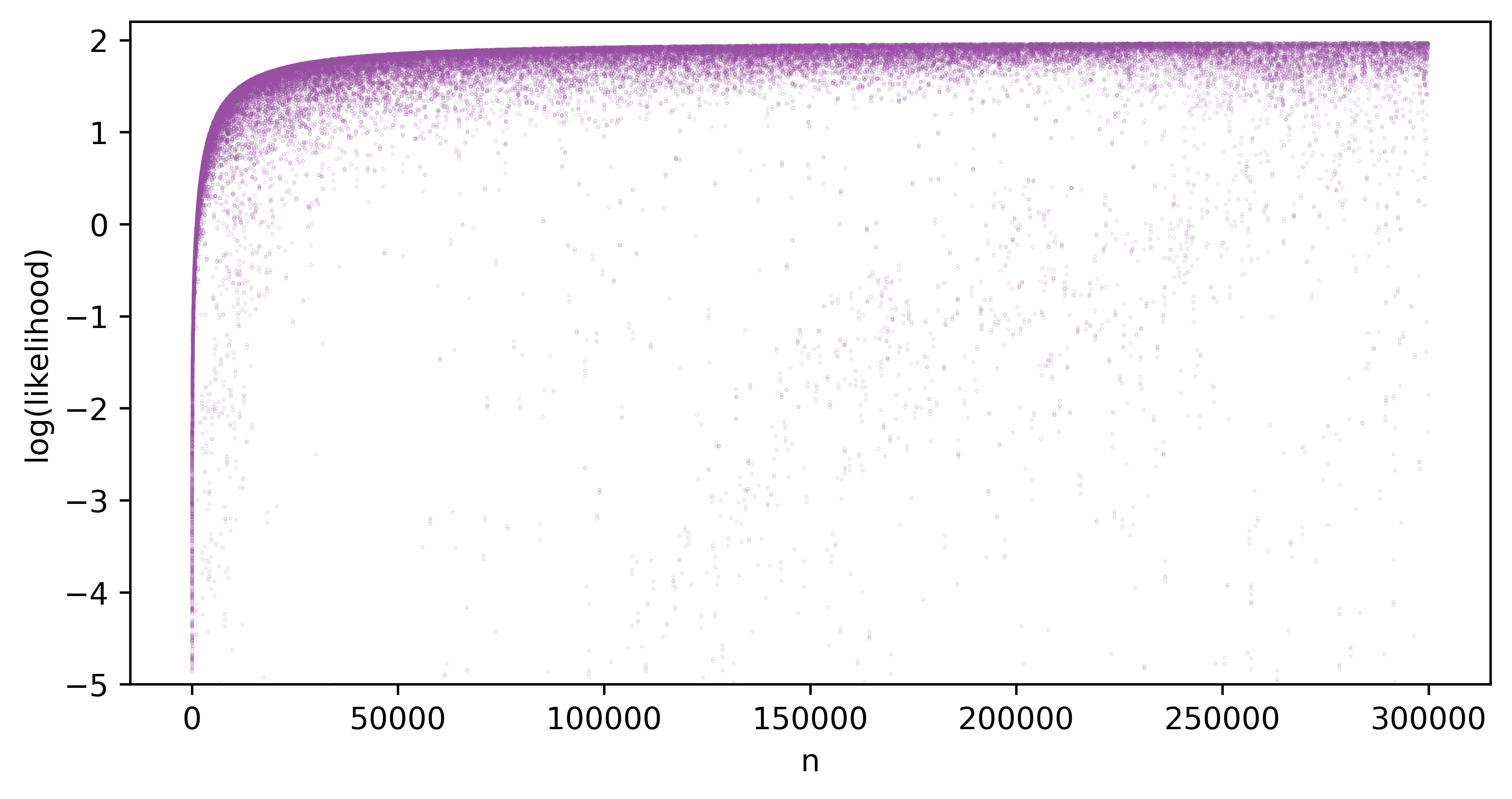}\end{center}\caption{The likelihood of a sample of datapoints, drawn from a sample of the
poseterior.}\label{fig:data_llike}\end{figure*}

Figure \ref{fig:data_llike} charts the log-likelihoods of this model.
The data is fit well, even for \(n > 220,000\) where the model uses a
sharp transition point instead of a smooth ``goose neck,'' and yet the
model systematically misrepresents the data for \(n < 100,000\). Even
so, it takes until approximately \(n < 1,100\) for the maximal
likelihood to drop below 1.

Note that the median times of Table \ref{tbl:small_n_fail} are all
within the 16/84th percentile of one another, and that the ordering does
not remain consistent. While this model does not give accurate
predictions for small \(n\), its assertion that execution times show
much more variance as \(n\) decreases is born out. If the spline
interpolation routine is not used on large lists, the performance
differences between all the variants are negligible.

The slope portions of the model show very confident fits. Surprisingly,
\texttt{newint\_\_noinv} has the fastest baseline of all the algorithms
while \texttt{newint\_\_vol} comes in second. A modified \emph{Numerical
Recipes}' algorithm, \texttt{splint\_one\_\_mul}, manages to have a
slightly better baseline than the original \texttt{newint} variant, and
the original \emph{Numerical Recipes} variant has the slowest baseline.
The baseline only represents the lowest possible execution time in
theory, however, and the Gamma distribution assigns it zero probability
of occuring. A fairer evaluation of execution times incorporates the
Gamma distribution, which not only adds an offset above the baseline but
also introduces variation in execution times.

\begin{lstlisting}[language=Python,numbers=none,xleftmargin=20pt,xrightmargin=5pt,belowskip=5pt,aboveskip=5pt]
plt.figure(num=None, figsize=(8, 4), dpi=600, facecolor='w', edgecolor='k')

ax = plt.subplot( 2, 1, 1 )

t = np.linspace( 5.5, 6.5, 1024 )

for i,alg in enumerate(algorithms):
    for idx,theta in enumerate(posteriors[i][0:100]):

        m_l3,m_mm,b_l3, t_l3_mm,w_s,w_p, mode,sigma_0,sigma_infty = theta

        shape, scale = gamma_param( mode, sigma_infty )

        if idx == 0:
            plt.plot( t, sps.gamma.pdf( t, shape, m_mm, scale ), color=colors[i], alpha=0.01, label=alg )
        else:
            plt.plot( t, sps.gamma.pdf( t, shape, m_mm, scale ), color=colors[i], alpha=0.01 )

leg = plt.legend( loc='best' )
for lh in leg.legendHandles:
    lh.set_alpha(1)

ax.set_title('n -> infty')
plt.xticks( [] )
plt.yticks( [] )

ax = plt.subplot( 2, 1, 2 )

n = 100000

for i,alg in enumerate(algorithms):
    for idx,theta in enumerate(posteriors[i][0:100]):

        m_l3,m_mm,b_l3, t_l3_mm,w_s,w_p, mode,sigma_0,sigma_infty = theta

        model = model_bilinear( theta, n )
        sigma     = model_noise_stdev( theta, n )
        shape, scale = gamma_param( mode, sigma )
        lower, upper = model_range_vargamma( theta, n )

        # the distribution itself
        if idx == 0:
            plt.plot( t, sps.gamma.pdf( t, shape, model, scale ), color=colors[i], alpha=0.01, label=alg )
        else:
            plt.plot( t, sps.gamma.pdf( t, shape, model, scale ), color=colors[i], alpha=0.01 )

ax.set_title('n = 100,000')
plt.xlabel( "nanoseconds / n" )
plt.yticks( [] )

plt.show()
\end{lstlisting}

\begin{figure*}[tbh]\begin{center}\adjustimage{max size={0.9\linewidth}{0.9\paperheight}}{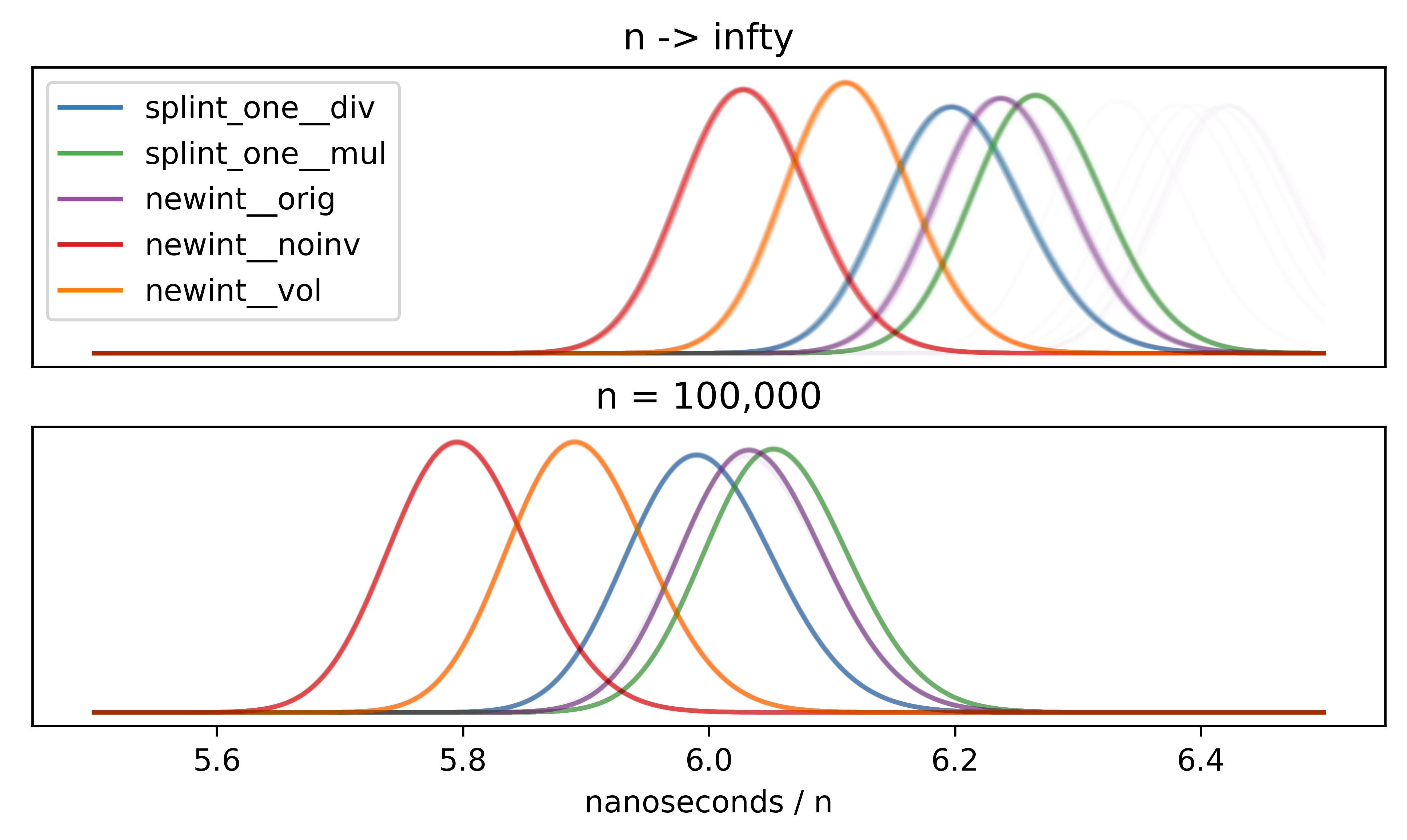}\end{center}\caption{Predicted execution times according to each algorithm's model, for large
\(n\). See text for details.}\label{fig:model_times}\end{figure*}

The upper half of Figure \ref{fig:model_times} is based on the model's
predition for \(n \to \infty\), while the lower half is based on
\(n = 100,000\). Both show that incorporating the noise component led to
the two \emph{Numerical Recipes} algorithms switching orders.

This provides consistent evidence for an unexpected conclusion, that
floating-point division does not carry the same penalty for the Intel
i7-7700k that it once did for older processors. Of the newer algorithms,
the fastest is the one which calculates the division last, without
loading it from memory, and converting a division into a multiplication
in \emph{Numerical Recipes}' algorithm resulted in a loss in
performance. More surprisingly, \texttt{newint\_\_vol}'s two additional
memory access result in a performance improvement over
\texttt{newint\_\_orig}. While this could be explained by the prior
observation that \texttt{gcc} will shift the division to occur near the
end of the algorithm, that code was compiled on \emph{Compiler Explorer}
with generic optimization flags. For the i7-7700k with
\texttt{-march=native}, \texttt{gcc} does not move the division, so the
only significant difference between \texttt{newint\_\_orig} and
\texttt{newint\_\_vol} is that the latter adds two \texttt{vmovsd}
instructions. The order of operations is slightly different, for
instance \texttt{newint\_\_orig} executes the division one instruction
earlier and interleaves some operations differently, but out-of-order
execution should eliminate any advantage that would provide.

\begin{lstlisting}[language=Python,numbers=none,xleftmargin=20pt,xrightmargin=5pt,belowskip=5pt,aboveskip=5pt]
arch_text = [ "AMD FX-4100",
              "Intel Xeon Gold 6148",
              "Intel Atom X5-Z8350",
              "Raspberry Pi 3 b+"
            ]

interp_range_bench = [ pd.read_csv("data/interp_range.fx-4100.clean.tsv", sep="\t" ),
                       pd.read_csv("data/interp_range.x6148.tsv", sep="\t" ),
                       pd.read_csv("data/interp_range.x5-z8350.tsv", sep="\t" ),
                       pd.read_csv("data/interp_range.pi-3b+.tsv", sep="\t" )
             ]

# control the y-axis range manually
y_limits = [ [12, 30],
             [12, 30],
             [35,100],
             [50,130],
              ]
\end{lstlisting}

\begin{lstlisting}[language=Python,numbers=none,xleftmargin=20pt,xrightmargin=5pt,belowskip=5pt,aboveskip=5pt]
plt.figure(num=None, figsize=(8, 6), dpi=600, facecolor='w', edgecolor='k')

for figure in range(4):

    ax = plt.subplot( 2, 2, figure+1 )

    subset = interp_range_bench[figure].sample( 64*1024 )

    for i,alg in enumerate(algorithms):

        # only include the middle execution
        mask = (subset['order'] & (0x7 << 3*2)) == ((i+1) << 3*2)
        plt.scatter( subset['n'][mask], (subset[alg]/subset['n'])[mask], alpha=0.3,
            marker='.', s=.1, color=colors[i], label="{}, round 3".format(alg) )

    plt.xticks( [0, 2.2e5, 1e6] )
    if figure > 1:
        plt.xlabel( "n" )
    if (figure & 1) == 0:
        plt.ylabel( "nanoseconds / n" )

    plt.ylim( y_limits[figure] )

    plt.tight_layout()
    ax.set_title(arch_text[figure])

plt.show()
\end{lstlisting}

\begin{figure*}[tbh]\begin{center}\adjustimage{max size={0.9\linewidth}{0.9\paperheight}}{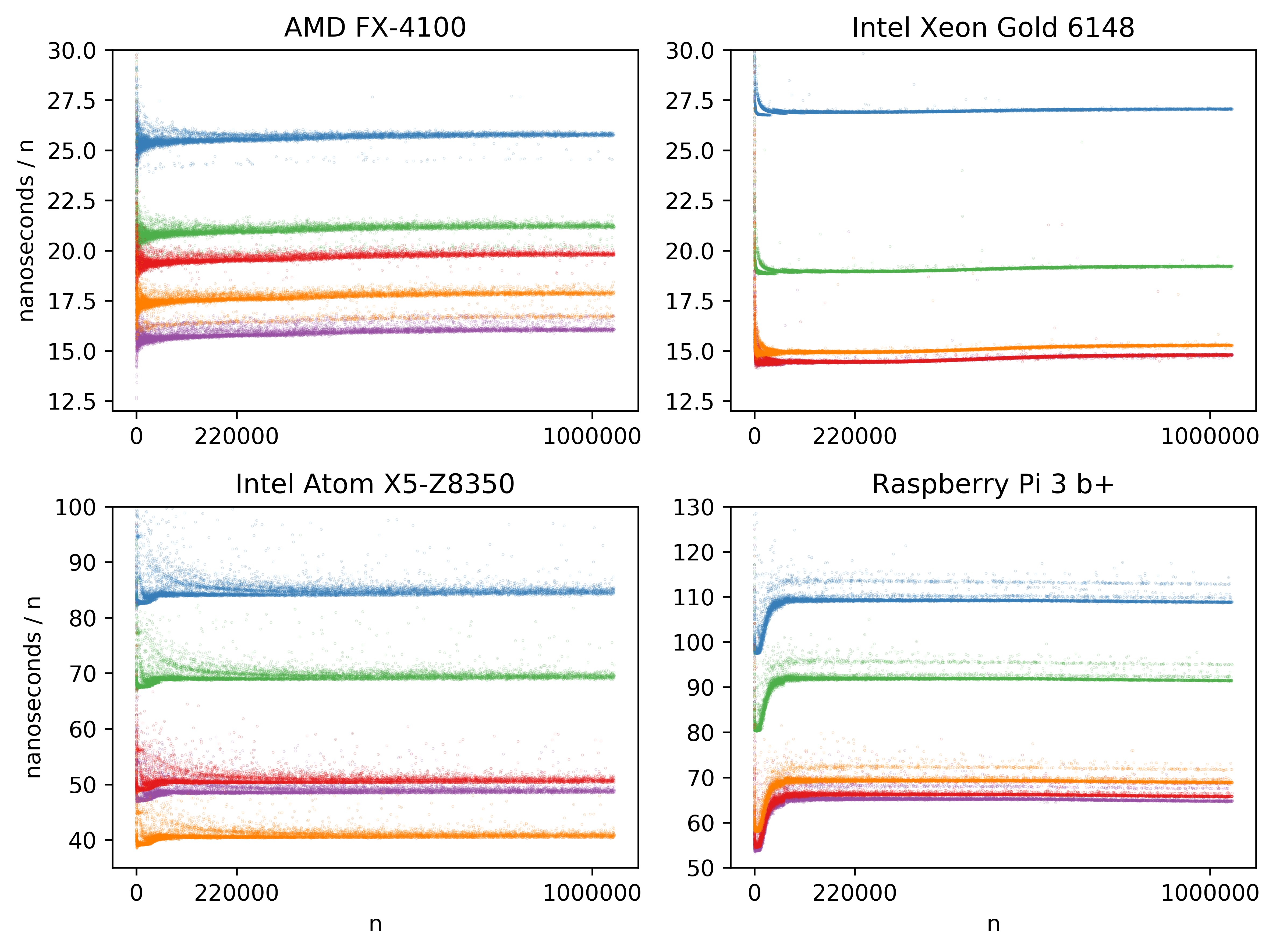}\end{center}\caption{Execution times for a variety of processor architectures. The colour
coding is the same as Figure ef\{fig:model\_times\}.}\label{fig:odd_procs}\end{figure*}

Running the same benchmark code on different processors provides further
evidence. Figure \ref{fig:odd_procs} shows the timing results from the
same code on an AMD FX-4100 (compiled with \texttt{gcc} 9.2.1), an Intel
Xeon Gold 6148 (Intel compiler 19.0.3.199), an Intel Atom x5-z8350
(\texttt{gcc} 9.2.1), and a Raspberry Pi 3 b+ (\texttt{gcc} 6.3.0). All
of them are either older, simpler, or more specialized than an i7-7700k,
and all of them provide evidence for division being a costly operation.
In all cases \emph{Numerical Recipes}' algorithm runs faster with one
division converted to a multiplication. On the Xeon Gold 6148
\texttt{newint\_\_orig} has the same performance as
\texttt{newint\_\_noinv}, while for all other microarchitectures the
former is consistently faster. The Atom x5-z8350 chart may suggest that
extra memory writes lead to a significant speed increase, but an
examination of the assembly code reveals that \texttt{gcc} has deferred
the division in \texttt{newint\_\_orig} to just before it's needed.
Unlike the other cases, the \texttt{volatile} hack is working as
intended.

None of these processors have the same cache layout as the i7-7700k, so
we'd expect their ``goose neck'' to be different or nonexistent. The
FX-4100 also has 8MiB of L3, so we'd expect performance to degrate at
roughly the same location as the i7-7700k, and we do observe that. But
while the latter's 256KiB L2 cache has been rendered useless long before
that point, leading to a flat line, the FX-4100's 2MiB L2 cache shared
between two cores is still gradually degrading and so it looks more like
a smooth curve than a ``goose neck.'' The x5-Z8350 lacks any L3 cache,
so the start of its ``goose neck'' instead marks when \(n\) occupies
half its 1MiB of L2 cache shared across two cores.

The Raspberry Pi 3 B+ also lacks an L3 cache, so its ``goose neck''
begins when half its shared 512KiB of L2 is occupied and ends when \(n\)
corresponds to roughly twice the size of L2. The Xeon Gold 6148 is very
similar, as its performance drop coincides with half its 1MiB-per-core
L2 cache. We expect to see a performance degredation when the dataset
exhausts half the 27.5 MiB L3 cache, but that occurs for
\(n > 6,040,234\).

Figure \ref{fig:odd_procs} also demonstrates there's less overlap
between the algorithms' performance for small \(n\). Switching from
\emph{Numerical Recipes}' code is much more likely to lead to a
performance increase for these microarchitectures.

More benchmarks on Intel processors help further demonstrate the
decreasing cost of division. On the
i5-3230m\footnote{Part of the Ivy Bridge microarchitecture, released in 2013}
and i7-4790\footnote{Haswell, 2014} processors, \emph{Numerical
Recipes}' \texttt{splint\_one} algorithm runs faster when the division
is converted to a multiplication. On an i3-6100\footnote{Skylake, 2015},
the conversion does not have a strong effect on performance. On the
i7-7700k\footnote{Kaby Lake, early 2017} and
i7-8700\footnote{Coffee Lake, late 2017}, the conversion causes a
decrease in performance. Interestingly, \texttt{newint\_\_noinv} has
better performance than \texttt{newint\_\_orig} on all the
microarchitectures listed in this paragraph. This is unlikely to be
explained by the extra memory fetch and multiplication in the latter
algorithm, as \texttt{newint\_\_vol} outperforms \texttt{newint\_\_orig}
for all but the i5-3230m despite adding two extra memory fetches.

\hypertarget{conclusion}{%
\section{Conclusion}\label{conclusion}}

Based on the results of these benchmarks, we can make a number of
suggestions for programmers hoping to implement this code.

The critical factor in deciding which algorithm to implement appear to
be the speed of the division instruction and the processor's ability to
organize out-of-order execution. On Intel processors with the Ivy Bridge
microarchitecture or better, out-of-order execution appears to favour
\texttt{newint\_\_noinv}, a variant of \texttt{newint} which eliminates
the \texttt{inv\_ba} variable in favour of division by \texttt{ba}. If
the architecture is Skylake or later, however, and the goal is to
interpolate evenly-spaced points across a curve with less than tens of
thousands of points, execution noise greatly reduces the performance
advantage for any one algorithm, including those from \emph{Numerical
Recipes}.

For processors with slow division, or those with weak or no ability to
reorder instruction execution, a straightforward implementation of
\texttt{newint} is likely to give the best performance, and to maintain
that performance for curves containing fewer than tens of thousands of
control points. CPUs designed for high-performance computation typically
fall into this category. One exception to look out for is if the
compiler shuffles the declaration of \texttt{inv\_ba} to occur
immediately before the \texttt{return} statement, which could impair
performance. Examine the assembly output to determine if this is
happening. If it is, either manually adjust the assembly code or declare
\texttt{inv\_ba} to be \texttt{volatile} to prevent the division from
migrating.

Surprisingly, on some processors the \texttt{volatile} hack leads to a
performance increase even though it introduces an extra memory store and
an extra fetch. This is unlikely to be faster than both a
straightforward implementation of \texttt{newint} and
\texttt{newint\_\_noinv}, but if performance is absolutely critical it
may be worth running some benchmarks to verify this.

As generating second derivatives is unlikely to be a performance
bottleneck, we did only a minimal comparison between \emph{Numerical
Recipes}' \texttt{spline} and our \texttt{new\_second\_derivative}.
Still, there is reason to expect that there will be little performance
difference between the two for small \(n\).

Our replacements for \texttt{spline} and \texttt{splint\_one} show only
minor divergences in accuracy, all of which can be explained by the
imprecision of floating-point operations. Other than the performance
differences outlined above, there should be no intrinsic obstacle to
replacing those \emph{Numerical Recipes} routines with ours.

In future, we would like to benchmark our code on more recent AMD
processors, as well as GPUs. The former are likely to run the
\texttt{newint\_\_noinv} variant faster than the original
\texttt{newint}, and the latter are likely to show the opposite
behaviour, but without running the experiment we cannot be sure. We are
uncomfortable with insisting that CPU benchmarks be single-threaded
while allowing GPU benchmarks unconstrained access to all hardware
execution units, so we would also like to perform some multithreaded CPU
benchmarks.

\bibliographystyle{unsrtnat}
\bibliography{Fast_Cubic_Spline_Evaluation_files/fast_cubic}

\end{document}